\documentclass[12pt]{article}
\usepackage{epsfig,epsf}

\newlength{\dinwidth}
\newlength{\dinmargin}
\def\lapproxeq{\lower .7ex\hbox{$\;\stackrel{\textstyle <}{\sim}\;$}}
\def\gapproxeq{\lower .7ex\hbox{$\;\stackrel{\textstyle >}{\sim}\;$}}

\def\be{\begin{equation}}
\def\ee{\end{equation}}
\def\bea{\begin{eqnarray}}
\def\eea{\end{eqnarray}}

\def\GeV{{\rm GeV}}

\begin{document}
\titlepage

\begin{flushright}
KEK--TH--1458\\
LTH 913\\
IPPP/11/18\\
DCPT/11/36\\
TU--883\\
15 June 2011\\
\end{flushright}

\vspace*{4cm}

\begin{center}
{\Large \bf $(g-2)_{\mu}$ and $\alpha(M_Z^2)$ re-evaluated using new
  precise data} 

\vspace*{1cm} {\sc Kaoru Hagiwara}$^{a}$, {\sc Ruofan Liao}$^b$, 
{\sc Alan D. Martin}$^c$, 
{\sc Daisuke Nomura}$^d$\\ and {\sc Thomas Teubner}$^{b}$

\vspace*{0.5cm}
$^a$ {\em KEK Theory Center and Sokendai, Tsukuba, 305-0801 Japan} \\
$^b$ {\em Department of Mathematical Sciences,\\
 University of Liverpool, Liverpool L69 3BX, U.K.}\\
$^c$ {\em Department of Physics and Institute for
Particle Physics Phenomenology,\\
University of Durham, Durham DH1 3LE, U.K.}\\
$^d$ {\em Department of Physics, Tohoku University, Sendai 980-8578, Japan}
\end{center}

\vspace*{1cm}

\begin{abstract}
We update our Standard Model predictions for $g-2$ of the muon and for
the hadronic contributions to the running of the QED coupling,
$\Delta\alpha^{(5)}_{\rm had}(M_Z^2)$. Particular emphasis is put on
recent changes in the hadronic contributions from new data in the
$2\pi$ channel and from the energy region just below $2$ GeV. In
particular, for the $e^+e^- \to \pi^+\pi^-$ contribution we include
the recent `radiative return' data from KLOE and BaBar. We also
include the recent BaBar data on other exclusive channels. We make a
detailed study of the effect of replacing the measurements of the
inclusive cross section, $\sigma(e^+e^- \to \mbox{hadrons})$, by the sum of
the exclusive channels in the energy interval $1.43 < \sqrt{s} < 2$
GeV, which includes a QCD sum-rule analysis of this energy region. 
Our favoured prediction for the muon anomalous magnetic moment is
$\,(g-2)/2 =  (11\,659\,182.8 \pm 4.9) \cdot 10^{-10}\,$ which is
$3.3\,\sigma$ below the present world-average measurement. We compare
our $g-2$ value with other recent calculations. Our prediction for the
QED coupling, obtained via $\,\Delta\alpha_{\rm had}^{(5)}(M_Z^2) =
\left( 276.26 \pm 1.38 \right)\cdot 10^{-4}\,$, is
$\,\alpha(M_Z^2)^{-1} = 128.944 \pm 0.019 \,$.
\end{abstract}
\section{Introduction}

The anomalous magnetic moment of the muon, $a_{\mu}=(g-2)_{\mu}/2$, provides
one of the strongest tests of the Standard Model (SM). The discrepancy
between the formidable measurement from BNL \cite{BNL} and the
theoretical prediction of $a_{\mu}$ is currently one of the few, if
not the only sign for physics beyond the SM (apart from neutrino
masses). At present, the discrepancy stands at about three standard
deviations, with comparable accuracy between experiment and
theory. Clearly, further progress is needed to scrutinize, and
possibly firmly establish, this discrepancy. On 
the theoretical side, this will require first and foremost the
improvement of the hadronic contributions which dominate the
uncertainty of the SM prediction. The hadronic
contributions are usually divided into the leading-order (LO) and
higher-order (HO) vacuum polarisation (VP) contributions, and the
so-called light-by-light scattering contributions, which are also
subleading in terms of the power counting in the coupling $\alpha$:
\begin{equation}
a_{\mu}^{\rm had} = a_{\mu}^{\rm had,\,LO\,VP} + a_{\mu}^{\rm had,\,HO\,
  VP} + a_{\mu}^{\rm had,\,l-by-l}\,.
\end{equation}
None of these contributions can be calculated reliably using
perturbative QCD (pQCD), as virtual photons with low $q^2$ dominate the loop
integrals. For the light-by-light contributions one relies on
model-calculations. A brief discussion of the status of these will be
given below when compiling our complete SM prediction of $g-2$. 
Fortunately, the situation is better under control for the VP
contributions which are large compared to the light-by-light
corrections; they can be predicted via dispersion integrals and
the experimentally measured hadronic cross section, $\sigma(e^+e^- \to
\gamma^* \to \mbox{hadrons})$. For $a_{\mu}^{\rm had,\,LO\,VP}$ the
relation reads 
\begin{equation}
a_{\mu}^{\rm had,\,LO\,VP} = \frac{1}{4\pi^3} \int_{m_{\pi}^2}^{\infty} {\rm d}s 
\, \sigma_{\rm had}^0(s) K(s)\,,
\label{eq:disp}
\end{equation}
where $\sigma_{\rm had}^0(s)$ is the undressed total hadronic cross
section (i.e. excluding VP corrections), and $K$ is a well known
kernel function given, e.g., by Eq.~(45) in the first reference of
\cite{HMNT03}. (Note that the normalisation of $K$ used here differs
by the factor $m_\mu^2/(3s)$.) At present a precision of about 1\% is
required  for the hadronic contributions. So it is mandatory to
combine, in the most reliable and consistent way, the best available
measurements from many experiments. Recently, several new measurements
have become available, both from `direct scan' experiments (like CMD-2
and SND at Novosibirsk's VEPP, and BES at Beijing's BEPC), and also
the radiative return data obtained in the recent analyses from KLOE
\cite{KLOE08,KLOE10} and BaBar \cite{BaBar2pi}. 

The main purpose of this work is to update our calculations
\cite{HMNT03,HMNT06} of the hadronic vacuum polarisation (HVP)
contributions to $g-2$ and $\Delta\alpha(M_Z^2)$. In section
\ref{sec:vp} we discuss recent changes in the HVP contributions to
$g-2$, detailing in subsection \ref{sec:vp2pi} the progress due to
radiative return analyses in the most important $2\pi$ channel. In
subsection \ref{sec:vpexcl} we study improvements in the important
energy region below $2$ GeV. In section \ref{sec:gm2comp} our updated
complete SM prediction of $g-2$ is given and compared to the BNL
measurement and other recent calculations of $g-2$. Our updated
evaluation of $\Delta\alpha(M_Z^2)$ is discussed in section
\ref{sec:delal}. Section \ref{sec:sum} contains our conclusions and
outlook. 

\section{Hadronic Vacuum Polarisation Contributions}
\label{sec:vp}
Since our last major update \cite{HMNT06} there have been significant
additions to the data input for $\sigma_{\rm had}^0$. Most
important, new precise data in the $2\pi$ channel have become
available, based on radiative return\footnote{For a review of this
  method, further references and recent results see \cite{MCSIGHAD}.}
analyses from KLOE \cite{KLOE08,KLOE10} and BaBar
\cite{BaBar2pi}. These data constitute a crucial check of the previous
measurements performed via the traditional method of energy scan by
adjusting the $e^+, e^-$ beams, and their impact will be discussed in
some detail below. Also, many new data sets in subleading channels
have been published, especially in the region just below $2$ GeV where
BaBar has measured many hadronic cross sections using radiative
return. In the following we will concentrate on these two energy
regions, before presenting our new results for $a_{\mu}^{\rm
  had,\,LO\,VP}$ and $a_{\mu}^{\rm had,\,HO\,VP}$. 

\subsection{Data combination and inclusion of data from radiative
  return experiments in the $2\pi$ channel} 
\label{sec:vp2pi}
In the dispersion integral (\ref{eq:disp}), the kernel function $K$
has the form $K = m_{\mu}^2/(3s) \cdot (0.4 \dots 1)$, where the
term in brackets stands for a function which increases monotonically
from $0.4$ to $1$. This results in a strong weighting towards low
energies and hence the dominance of the $\rho \to 2\pi$ channel, which
makes up more than 70\% of $a_{\mu}^{\rm had}$. In recent years this
channel has been  measured very precisely via the direct scan method
(adjusting the $e^+, e^-$ beam energy) by the CMD-2 and SND
experiments at Novosibirsk, see e.g. \cite{Ignatov} for a brief review
of their results. From 2005 onwards, analyses based on the method of
radiative return have also become available. The first analysis of this
kind for the $2\pi$ channel was published by KLOE \cite{KLOE05}, and
their $2\pi$ distribution agreed fairly well with the measurements
from CMD-2 and SND, although there were some shape-differences
compared to the CMD-2 and SND data.

In \cite{HMNT06} we had already taken into account these KLOE data
which were published in 2005. However, the slight difference in shape
had made it difficult to combine the KLOE data with all the other
$2\pi$ data sets on a bin-to-bin basis, and we had hence combined the
KLOE data only {\em after} the dispersion integration, see the
detailed discussion in \cite{HMNT06}. However, a combination of all
data sets in one spectral function and on the same footing, i.e.\ {\em
  before integration}, is clearly preferable, as possible
inconsistencies between different data sets will lead to a
well-defined error estimate in the data compilation and hence in
$a_{\mu}^{\rm had}$. 
Therefore, in the current analysis, we have included all data sets in
the same way, by performing a $\chi^2_{\rm min}$ fit similar to our
original procedure, as described in detail in \cite{HMNT03}. This
includes the new $2\pi(\gamma)$ data from the radiative return
analyses of KLOE \cite{KLOE08,KLOE10}\footnote{Note that the KLOE08
  data \cite{KLOE08} supersede their earlier analysis from 2005
  \cite{KLOE05}, which is therefore discarded in our new compilation.}
and BaBar \cite{BaBar2pi}. 

For the data sets from BaBar and KLOE full covariance matrices for
statistical and systematic errors are available. To take these into
account consistently, our original $\chi^2$ function has to be
modified to include additional contributions from non-diagonal
elements of the covariance matrices for the single data sets. The
function now reads 
\begin{eqnarray}
\chi^2(R_m,f_k) & = & \sum^{N_{\rm exp}}_{k=1}
 \left(\frac{1-f_k}{{\rm d}f_k}\right)^2 + \nonumber \\
&  & \left\{
  \sum^{N_{\rm clu}}_{m=1}\sum^{N_{(k,m)}}_{i=1}
  \left(\frac{R^{\,(k,m)}_i-f_kR_m}{{\rm d}R^{\,(k,m)}_i}\right)^2
     \right\}_{\rm w/out\ cov.\, mat.} + \nonumber \\
&  & 
  \sum^{N_{\rm clu}}_{m,n=1}\sum^{N_{(k,m)}}_{i=1}\sum^{N_{(k,n)}}_{j=1}  
  \left(R^{\,(k,m)}_i-f_kR_m\right)C^{-1}(m_i,n_j)
  \left(R^{\,(k,n)}_j-f_kR_n\right) \,,
\label{eq:chisq}
\end{eqnarray}
where the notation of \cite{HMNT03} has been adopted.
The final term of the $\chi^2$ function (\ref{eq:chisq}) allows the
inclusion of the full covariance matrices as given for individual data
sets; here $C^{-1}$ denotes the inverse of the matrix $C$ which is
the sum of the statistical and systematic covariance matrices of the
data set contributing to the specific term in the triple
sum.\footnote{Note that all the common systematic errors are already
  taken into account via ${\rm d}f_k$ in the first term of the $\chi^2$
  function and therefore subtracted from the matrices $C$.} 
The subscript `w/out cov. mat.' in the penultimate term indicates that
contributions taken into account via a full covariance matrix are
excluded to avoid double counting. 
For example, BaBar gives experimental uncertainties in the 
form of a covariance matrix which contains ``non-diagonal statistical
errors''. In such cases, we take the statistical and systematic
contributions to the $\chi^2$ function into account in the third term
(and the common systematic error in the first term), but do not
include them in the second term. For those experiments whose
experimental uncertainties can be considered as an overall systematic
and (diagonal) statistical errors, we take them into account in
the first and second term, but not in the third term. 
As in our previous works, this non-linear $\chi^2$ function
is minimised numerically, returning the fitted `cluster' (energy bin)
mean values $R_m$ and renormalisation factors $f_k$, in addition to
the full covariance matrix of the fit which is used for our error
estimate in the prediction of $a_{\mu}^{\rm had}$ and
$\Delta\alpha$.\footnote{We do not attempt to estimate correlations
  between different experiments or between different channels from the
  same experiment. While such correlations certainly exist,
  e.g. through common Monte Carlo codes or luminosity uncertainties,
  it is not clear to us how to quantify these effects in a
  well-defined way. Davier {\it et al.} \cite{Davier:2010nc} claim
  significant effects, but note that their error from the exclusive
  channels is very similar to ours (see
  Table~\ref{tab:comparisonDavieretal} below).} 
\begin{table}[htb]
\begin{center}
\begin{tabular}{l|c|c|c|c}
\hline
Channel & global $\chi^2_{\rm min}/{\rm d.o.f.}$ & globally infl.\
err. & locally infl.\ err. & `global -- local' \\ 
\hline
$2\pi$                   & $1.4$ & $3.06$ & $3.09$ & $-0.03$ \\
$3\pi$                   & $3.0$ & $1.08$ & $0.99$ & $+0.10$ \\
$4\pi(2\pi^0)$           & $1.3$ & $1.19$ & $1.26$ & $-0.07$ \\
$4\pi(\mbox{no}\,\pi^0)$ & $1.7$ & $0.49$ & $0.47$ & $+0.02$ \\
$K^+K^-$                 & $1.9$ & $0.57$ & $0.46$ & $+0.11$ \\
$K^0_SK^0_L$             & $0.8$ & $0.16$ & $0.16$ & $-0.003$ \\
$5\pi(1\pi^0)$           & $1.2$ & $0.09$ & $0.09$ & $0$ \\
$6\pi(2\pi^0)$           & $4.0$ & $0.39$ & $0.24$ & $+0.16$ \\
\hline
\end{tabular}
\end{center} 
\vspace{-2mm}
\caption{Global $\chi^2_{\rm min}/{\rm d.o.f.}$, globally and locally
  inflated error of $a_{\mu}$ and their difference for several
  channels. (Range of integration from threshold to $2$ GeV.) The five
  and six pion channels are used as input for our updated isospin
  analysis (see below).} 
\label{tab:globallocalchisq}
\end{table}
In \cite{HMNT03,HMNT06} we had applied error inflation
by the global $\sqrt{\chi^2_{\rm min}/{\rm d.o.f.}}$ if bigger than
one, indicating that the data are not compatible within
errors. However, this may be considered a crude approximation,
especially in the case of $a_{\mu}$ where the weighting in the
dispersion integral is far from flat. Therefore we have improved the
error estimate by employing local $\chi^2$ inflation. This is achieved
by calculating, for the preferred binning (judged by the global
$\chi^2_{\rm min}/{\rm d.o.f.}$ in the respective channel), a {\em local}
$\chi^2$ {\em in each cluster}. This is then used to inflate the error 
locally when calculating the error by use of the full covariance
matrix. 
The errors obtained in this way vary slightly from the ones based on
global $\chi^2$ inflation and are smaller in most channels. This is
indicated in Table~\ref{tab:globallocalchisq}, where, for the most
important channels, the global $\chi^2_{\rm min}/{\rm d.o.f.}$ is
displayed together with the errors based on global and local error
inflation and their difference. For the numbers used in this analysis,
we will use local inflation whenever applicable. 

\begin{figure}[htb]
\begin{center}
\psfig{file=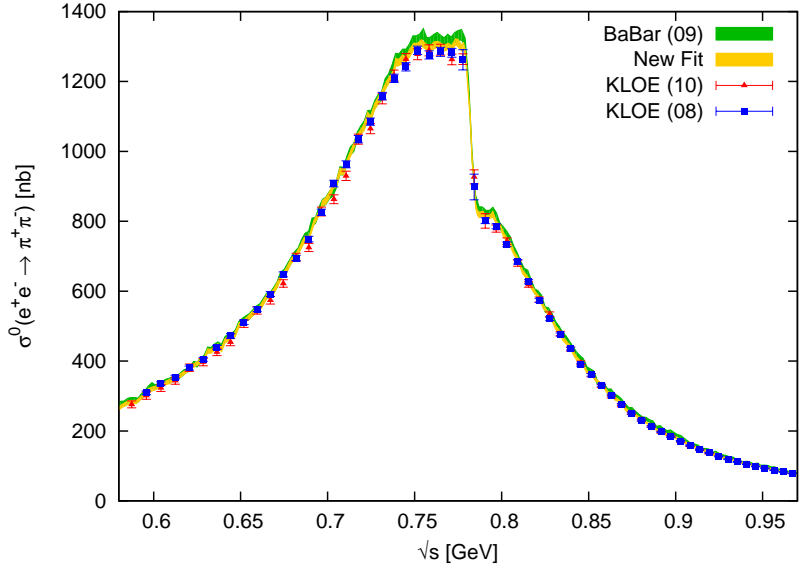,width=11cm,angle=0}
\end{center}
\vspace{-0.6cm}
\caption{Fit with all data in the $2\pi$ channel: light (yellow)
  band. Radiative return data from BaBar \cite{BaBar2pi} are shown by
  the darker (green) band, whereas the KLOE \cite{KLOE08,KLOE10} data
  are displayed by the markers as indicated in the plot.}  
  \label{fig:2pidataall}
\end{figure}
\begin{figure}[htb]
\begin{center}
\psfig{file=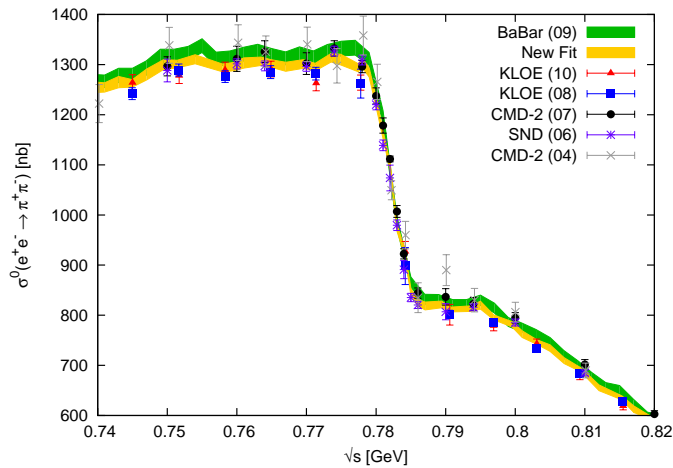,width=9cm,angle=0}
\end{center}
\vspace{-0.6cm}
\caption{Same as Fig.~\ref{fig:2pidataall}, but `zoomed' in for the
  $\rho-\omega$ interference region. In addition to the radiative
  return data, also the important data from CMD-2
  \cite{CMD2old,CMD2new} and SND \cite{SNDre} are displayed.} 
  \label{fig:2pidatarhoom}
\end{figure}
\begin{figure}[htb]
\begin{center}
\psfig{file=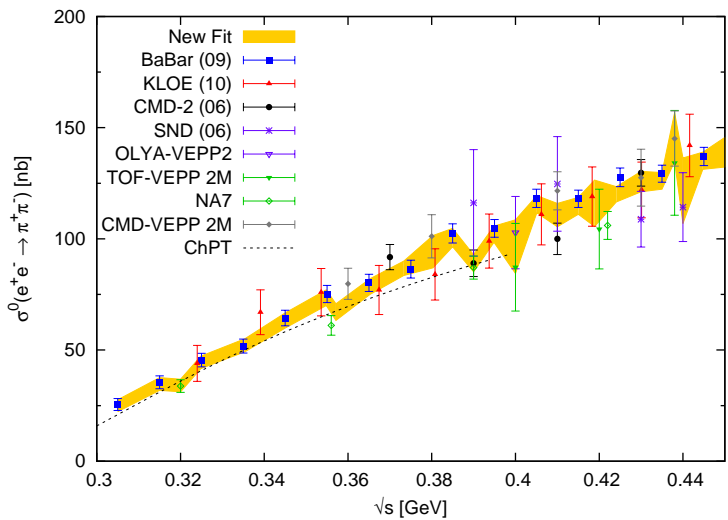,width=10cm,angle=0}
\end{center}
\vspace{-0.6cm}
\caption{Low energy tail of the $\rho \to 2\pi$ channel: the light
  (yellow) band shows the result of our fit using all data, whereas the
  markers display the actual data points as indicated in the plot. The
  dashed line represents the prediction of chiral perturbation theory
  (labelled ChPT) used for the lowest energies from threshold to the
  first BaBar point at $0.305$ GeV.} 
  \label{fig:2pidatalow}
\end{figure}
\begin{figure}[htb]
\begin{center}
\psfig{file=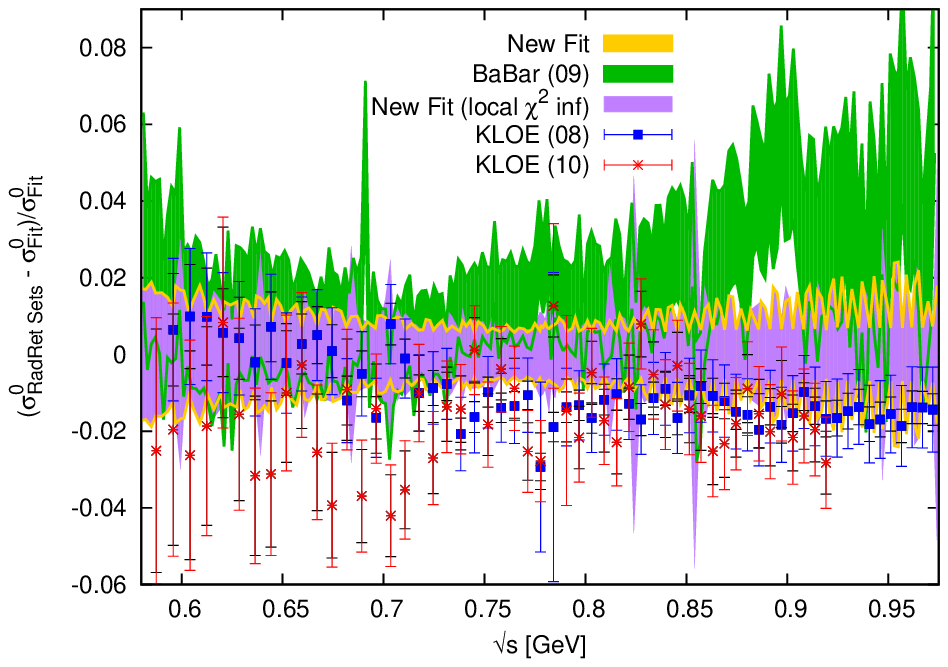,width=11cm,angle=0}
\end{center}
\vspace{-0.6cm}
\caption{Normalised difference between the data sets based on
  radiative return from KLOE \cite{KLOE08,KLOE10} and BaBar
  \cite{BaBar2pi} and the fit of all data in the $2\pi$ channel, as
  indicated on the plot. The (lilac) band symmetric around zero
  represents the error band of the fit given by the diagonal elements
  of the fit's covariance matrix, with local error inflation as
  explained in the text, whereas the light (yellow) band indicates the
  error band of the fit without inflation.} 
  \label{fig:radretvsnewfit}
\end{figure}
The role of the radiative return data from KLOE and BaBar in the new
fit is demonstrated in Fig.~\ref{fig:2pidataall} in the $\rho$ region
from about 0.6 to 0.95 GeV. The new data from
BaBar \cite{BaBar2pi} are represented by the darker (green) band,
whereas the data from KLOE \cite{KLOE08,KLOE10} are displayed by the
markers as indicated on the plot. The light (yellow) band is the
result of the fit of all combined $2\pi$ data, i.e. the data as used
in \cite{HMNT06} together with the new data from KLOE and BaBar. 
Figure~\ref{fig:2pidatarhoom} shows a zoom into the peak region with
the $\rho - \omega$ interference and also displays important data from
the experiments CMD-2 \cite{CMD2old,CMD2new} and SND
\cite{SNDre}. Figure~\ref{fig:2pidatalow} displays the low energy
region close to threshold, a region previously only sparsely populated
by data and where BaBar has added very valuable information.
It is clear already from these figures that the KLOE 
data\footnote{The KLOE08 data are in very good agreement with those of
  the independent KLOE10 analysis.} are lower than the BaBar data, and
the fit interpolates between both. Note that despite this tension the
fit quality is rather good, with a (global) $\chi^2_{\rm min}/{\rm
  d.o.f.} = 1.4$. 
The difference between the KLOE and BaBar data and the full fit to all
data is exemplified in Fig.~\ref{fig:radretvsnewfit}, where normalised
differences between the sets from radiative return and the fit of all
data are displayed as indicated on the plot. 

\begin{figure}[htb]
\begin{center}
\hspace{10mm}\psfig{file=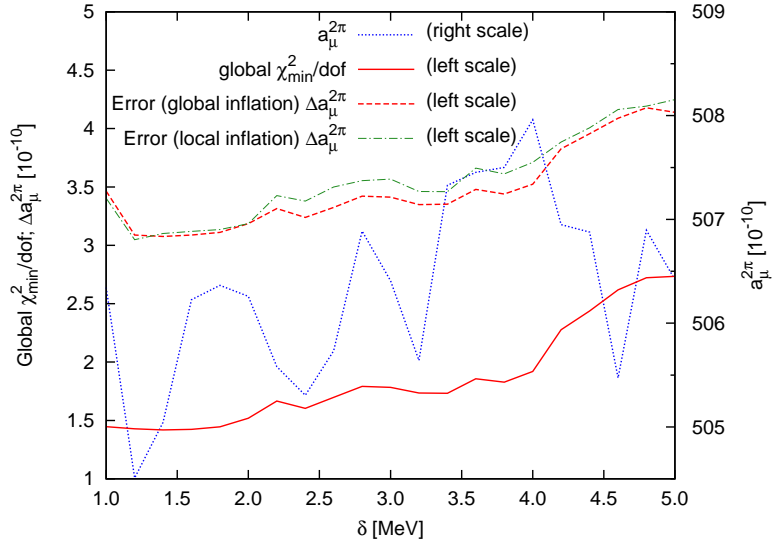,width=10.5cm,angle=0}
\end{center}
\vspace{-0.6cm}
\caption{Dependence of the global $\chi^2_{\rm min}/{\rm d.o.f.}$ (solid red
  line, left scale), the globally inflated error $\Delta
  a_{\mu}^{2\pi}\cdot 10^{10}$ (dashed red line, left scale) and the
  mean value $a_{\mu}^{2\pi}\cdot 10^{10}$ (dotted blue, right scale)
  on the choice of the cluster size parameter $\delta$. The
  dash-dotted green line indicates $\Delta a_{\mu}^{2\pi}\cdot
  10^{10}$ with local error inflation. Recall $a_{\mu}^{2\pi}$ is the
  $2\pi$ contribution in the range $0.305 < \sqrt{s} < 2$ GeV.} 
  \label{fig:clusteringscan}
\end{figure}
One may ask how stable the fit is with respect to different energy
binnings or assumptions on the cross section. There are now a very
large number of data points\footnote{We use 879 data points in the
  $2\pi$ channel.} coming from several different experiments. Thus an
extremely fine binning is possible and hence differences, or biases,
due to varying the underlying model for the cross section are
negligible.\footnote{As we have checked and discussed in
  \cite{HMNT03}, our simple assumption of a piecewise constant cross
  section in the energy bin and simple trapezoidal integration are well
  justified.} However, there is a remaining dependence on the way the
data are binned. For the current analysis, we have further refined our
algorithm to `cluster' the data in energy bins with a given target
cluster size $\delta$. Compared to the original algorithm used in
\cite{HMNT03}, we now allow for more flexible bin-sizes in
regions of very dense or sparse data.\footnote{This adaptive
  clustering avoids individual bins containing too many data points,
  and prevents points very close to each other from being forced into
  different bins, as could happen with a fixed cluster size $\delta$.}
The dependence on the target cluster size $\delta$ is displayed in
Fig.~\ref{fig:clusteringscan} for a range of $1\,{\rm MeV}< \delta <
5\,{\rm MeV}$: the solid (red) line gives $\chi^2_{\rm min}/{\rm
  d.o.f.}$, the dashed (red) line the error of the resulting $2\pi$
contribution to $a_{\mu}$ (after error inflation and in units of
$10^{-10}$) and the dotted (blue) line shows the variation of the mean
value of $a_{\mu}^{2\pi}$ (again in units of $10^{-10}$). This
demonstrates that, (i) data clustering with cluster sizes $\delta$ up
to about $2$ MeV is advantageous as the fit quality is maintained and
the error reduced, (ii) larger clustering is disfavoured by an
increase in $\chi^2_{\rm min}/{\rm d.o.f.}$ which overcompensates any
possible additional gain of the error before error inflation, (iii)
the mean value of the resulting $a_{\mu}^{2\pi}$ varies by about
$1\,\sigma$, depending on the particular choice of the clustering. We
have chosen $\delta = 1.5$ MeV which leads to the best fit quality,
the smallest error and a mean value of $a_{\mu}^{2\pi}$ right in the
middle of the band of possible results for various clustering
choices. With this we obtain 
\begin{equation}
a_{\mu}^{2\pi}(0.32 - 2\ {\rm GeV}) = (504.23 \pm 2.97) \cdot
10^{-10}
\label{eq:amu2pi}
\end{equation}
for the energy range $0.32 < \sqrt{s} < 2$ GeV (note
Fig.~\ref{fig:clusteringscan} is for the different interval $0.305 <
\sqrt{s} < 2$ GeV). 
This result is to be compared to the result without the radiative return
data (337 points from BaBar, 60 from KLOE08 and 75 from KLOE10), for
which we would get 
$$
a_{\mu}^{2\pi,\,{\rm w/out\ Rad.\,Ret.}}(0.32 - 2\ {\rm GeV}) =
(501.26 \pm 4.48) \cdot 10^{-10}\,,
$$
if the same small clustering size $\delta = 1.5$ MeV is used. The
inclusion of the data from radiative return leads therefore to a
considerable pull-up of our previous result. Even more, without the
many data points from BaBar and KLOE the fit would prefer the much
bigger cluster size of $4.2$ MeV (as used in \cite{HMNT06}), for which
we would get 
\begin{equation}
a_{\mu}^{2\pi,\,{\rm w/out\ Rad.\,Ret.}}(0.32 - 2\ {\rm GeV}) =
(498.65 \pm 3.28) \cdot 10^{-10}\,.
\label{eq:amu2piwoutradret}
\end{equation}
The observed significant pull-up of $a_{\mu}^{2\pi}$ is expected from
Fig.~\ref{fig:radretvsnewfit}, and similar effects from the inclusion
of radiative return data have been reported in \cite{DavierBaBar}.
Given the small error of the new data based on radiative return, the
gain in accuracy seems modest. However, this is a consequence of the
tension between the different data sets as discussed above. 

It is interesting to compare $a_{\mu}^{2\pi}$ in the restricted energy
range where the KLOE (and BaBar) data overlap with the other data.
In the range $0.5958 < \sqrt{s} < 0.9192$ GeV the $a_{\mu}^{2\pi}$ (in units of
$10^{-10}$) using the KLOE08 data gives $376.3 \pm 3.4$, in fair
agreement with KLOE10 for which we get $373.4 \pm 3.3$, whereas using
the BaBar data results in $384.4 \pm 2.8$. In the same range and using
our original choice of $\delta = 4.2$ MeV, the integral over the
compilation of all other (excluding radiative return) data yields
$376.0 \pm 2.6$. A weighted average of these numbers would result in a
prediction of $377.9 \pm 1.5$. In comparison, combining all sets
before integration in this energy range gives $380.0 \pm 2.2$. As
expected from the discussion of the full energy range, this result is
considerably higher; and the larger error is a direct consequence of
the tension between the various sets, which would be partly masked if
the combination is done after integration. Therefore, for our updated
SM prediction of $g-2$ given below, we will use all $2\pi$ data in the
same way, i.e. combine them before integration and use
Eq.~(\ref{eq:amu2pi}) for the $2\pi$ contribution. 

Note that for our compilation of the hadronic cross section, and in
particular for the $2\pi$ channel, we do not use spectral function
data from hadronic $\tau$ decays, but prefer to entirely rely on 
$e^+e^- \to \mbox{hadrons}$ cross sections. Considerable effort has been
invested in model-based estimates of the isospin breaking corrections
needed to translate $\tau \to \nu_\tau W \to \nu_{\tau}\pi\pi^0$ ($I =
1$) spectral function data into the $e^+e^- \to \gamma^* \to
\pi^+\pi^-$ cross section.\footnote{The $I=0$ part, which is not
  present in the $\tau$ decays, contributes about 25\% to
  $a_{\mu}^{2\pi}$ and has to be estimated from $e^+e^-$ data, which
  contain the coherent sum of both $I=1$ and $I=0$ contributions. (See
  \cite{WolfeMaltman} for a recent evaluation of $\rho-\omega$ mixing
  contributions and its model dependence.)} However, even after
application of the known isospin breaking corrections, differences
between the two quantities remain. In their recent works
\cite{Davier:2009ag,Davier:2010nc}, 
Davier {\it et al.} have found, compared to their earlier studies, a
diminished, though still sizeable discrepancy between the isospin
rotated $\tau$ data from ALEPH, OPAL, CLEO and BELLE, and the $e^+e^-$
data. This is at variance with results from Benayoun {\it et al.}, who
found agreement of the $\tau$ with the $e^+e^-$ data in an analysis
based on hidden local symmetry and dynamical $(\rho, \omega, \phi)$
mixing \cite{Benayoun}. 
In a very recent work, Jegerlehner and Szafron \cite{JS} study the
effect of $\rho-\gamma$ mixing, which is absent in the $\tau$ spectral
function, but can be estimated from the $e^+e^-$ data. They find that
after including these effects the $\tau$ spectral function data
confirm the $e^+e^-$ data and are very close to them, but lead only to
a marginal improvement in the accuracy of $a_{\mu}^{\rm had}$.
A detailed discussion of these issues goes beyond the scope of this
work. However it is fair to say that there are still many open
questions. Given the limited understanding of the hadronic dynamics
and the resulting model dependence of the isospin breaking
corrections, and given the small influence of the $\tau$ data, we, at
present, do not include them for the best possible prediction of
$g-2$. 

\subsection{The energy region below $2$ GeV}
\label{sec:vpexcl}
Another region, where important changes have occurred through new
data, is the energy region between $1.43$ and $2$ GeV. This region is
particularly difficult, as, compared to lower energies, many more
multi-hadron exclusive final states are open and have to be included
to obtain an accurate prediction of $\sigma_{\rm had}$. However, this
energy range has been inaccessible for recent experiments
measuring low energy hadronic cross sections, especially CMD-2 and SND
operating at VEPP-2M in Novosibirsk.\footnote{This will change in the
  near future, with the experiments CMD-3 and SND already taking data
  at the upgraded VEPP-2000 collider, see \cite{VEPP-2000}.} Hence the
quality of the available data 
was not very good, with some channels hardly constrained at all. 
Alternatively, one can rely on inclusive $R$ measurements, but also for
these only rather old and not very precise data are available, with
poorly constrained systematics and limited statistics. The situation
has recently changed significantly, with BaBar measuring, again
through the method of radiative return, many channels with higher
accuracy than earlier experiments. These include new measurements of
the channels $2\pi^+2\pi^-$ \cite{Babar4pi}, $K^+K^-\pi^0$,
$K^0_SK\pi$ \cite{BaBar_1}, $2\pi^+2\pi^-\pi^0$, $K^+K^-\pi^+\pi^-\pi^0$,
$2\pi^+2\pi^-\eta$ \cite{BaBar_2}, $2\pi^+2\pi^-2\pi^0$
\cite{BaBar_3}, all of which we use for this updated
analysis.\footnote{A partial update of our analysis in this region has
  already been reported at the PhiPsi09 conference \cite{HLMNT1}.}
\begin{figure}[htb]
\begin{center}
\psfig{file=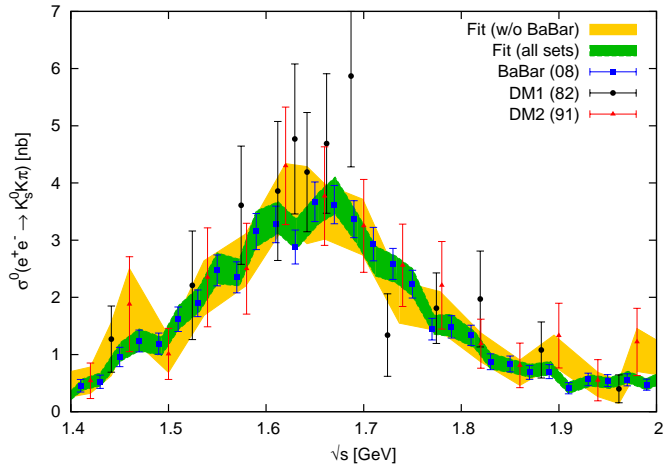,width=9cm,angle=0}
\end{center}
\vspace{-0.6cm}
\caption{$K_S^0 K \pi$ channel with improvement due to
  recent BaBar data \cite{BaBar_1} from radiative return.}
  \label{fig:ch-05}
\end{figure}
\begin{figure}[htb]
\begin{center}
\psfig{file=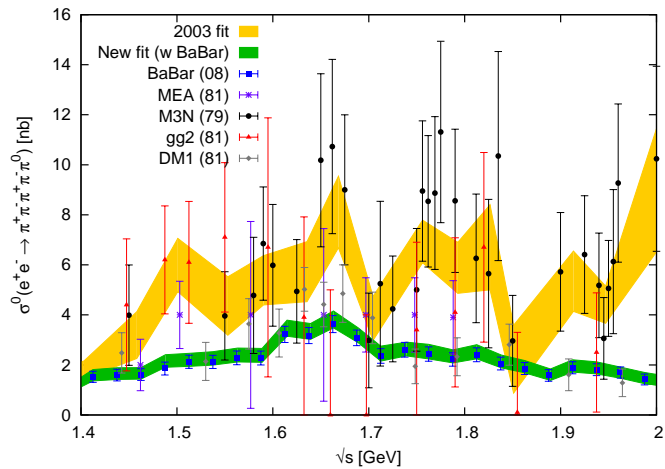,width=9cm,angle=0}
\end{center}
\vspace{-0.6cm}
\caption{$2\pi^+ 2\pi^- \pi^0$ channel with data from BaBar
  \cite{BaBar_3}. The data from M3N are not taken into account in the
  new fit.} 
  \label{fig:ch-14}
\end{figure}

\begin{figure}[htb]
\begin{center}
\psfig{file=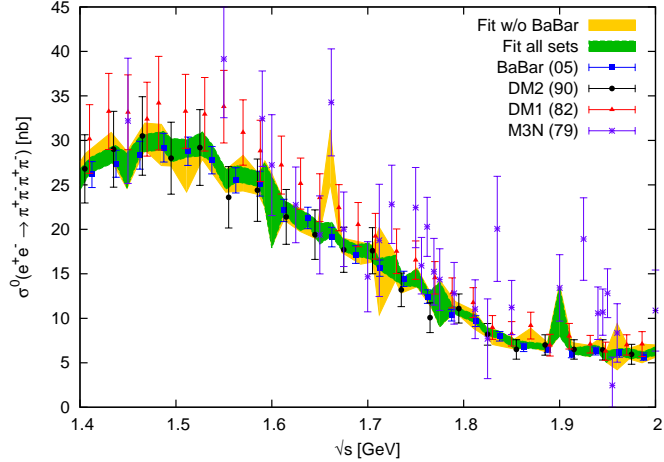,width=9cm,angle=0}
\end{center}
\vspace{-0.6cm}
\caption{$2\pi^+ 2\pi^-$ channel with data from BaBar \cite{Babar4pi}.}
\label{fig:ch-12}
\end{figure}

\begin{figure}[htb]
\begin{center}
\psfig{file=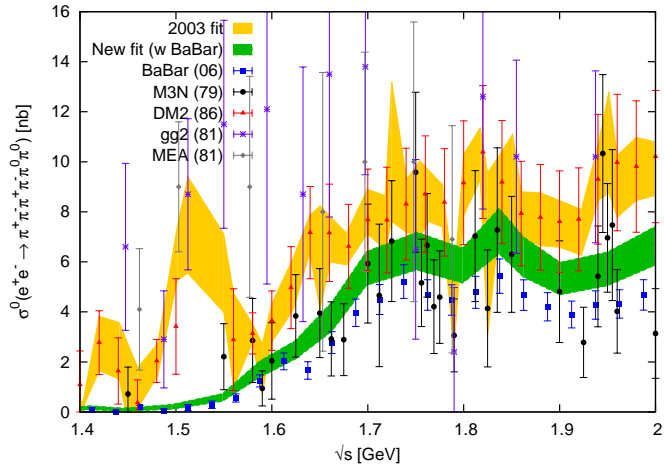,width=9cm,angle=0}
\end{center}
\vspace{-0.6cm}
\caption{$2\pi^+ 2\pi^- 2\pi^0$ channel with data from BaBar
  \cite{BaBar_3}. The data from M3N and DM2 are not taken into account
in the new fit.}  
\label{fig:ch-15}
\end{figure}

Figures~\ref{fig:ch-05} -- \ref{fig:ch-15} exemplify the influence of
the new BaBar data in the compilations for the subleading channels
$K_S^0 K \pi$, $2\pi^+ 2\pi^- \pi^0$, $2\pi^+ 2\pi^-$ and $2\pi^+
2\pi^- 2\pi^0$. In all 
these figures the light (yellow) bands indicate our older compilations
without the recent BaBar data, whereas the darker (green) bands show
the results of our new fits. The most important data are displayed
by markers as indicated in the plots. (For full references to older
data see \cite{HMNT03,HMNT06}.) Note that the new data from BaBar now
allow us to discard older, not very precise data from M3N (in the
$5\pi$ and $6\pi$ channels) and from DM2 ($5\pi$), which were only
available from theses and which are not compatible with the new
measurements. In the $2\pi^+ 2\pi^- \pi^0$ channel we have also
included data from DM1 \cite{DM1}, with properly added non-resonant
contributions. All the channels shown here are now somewhat dominated
by the radiative return measurements from BaBar. Only in the $2\pi^+
2\pi^- 2\pi^0$ channel do other data still influence the fit in a
significant way through a slight adjustment via the fitted
renormalisation factors $f_k$ of the $\chi^2$ minimisation. 

As is clear from figures \ref{fig:ch-05} -- \ref{fig:ch-15}, the new
data are not always in agreement with earlier measurements. In case of
disagreement of different data sets, the fit has a bad (global)
$\chi^2_{\rm min}/{\rm d.o.f.}$. As done throughout our analysis, in
such cases the error of the channel's contribution to $a_{\mu}$ (and
$\Delta\alpha$) is scaled up by applying local error inflation through
$\chi^2$ values calculated in each cluster of the data compilation. 
This happens, for example, in the $2\pi^+ 2\pi^- 2\pi^0$ channel where
$\chi^2_{\rm min}/{\rm d.o.f.} = 2.7$, but see
Table~\ref{tab:globallocalchisq} for details.

\begin{table}[htb]
\begin{center}
\begin{tabular}{c|c|cl}
\hline
Channel & This work & HMNT (03) \cite{HMNT03} & approach used
          in \cite{HMNT03} \\
\hline
\rule{0pt}{2.75ex}
$K\bar K\pi$        & $2.77\pm0.15$ & $2.58\pm0.26$ &
 sum of $K^+K^-\pi^0,\,K^0_{S,L}K\pi,\,K^0_SK^0_L\pi^0$ \\
$K\bar K2\pi$       & $3.31\pm0.58$ & $3.63\pm1.24$ & 
 different relation based on channel $K_S^0X$ \\
$K\bar K3\pi$       & $0.08\pm0.04$ & -             & 
 not accounted for \\
$\omega(\rightarrow\pi^0\gamma)K\bar K$ 
                    & $0.01\pm0.00$ & -             &
 not accounted for \\[1ex]
$2\pi^+2\pi^-\pi^0\,({\rm no}\,\eta)$
                    & $1.20\pm0.10$ & $2.85\pm0.25$ &
 purely data-based incl.\ $\eta$ \\
$\pi^+\pi^-3\pi^0\,({\rm no}\,\eta)$
                    & $0.60\pm0.05$ & $1.19\pm0.33$ & 
 based only on M3N data \\
$\omega(\rightarrow\pi^0\gamma)2\pi$
                    & $0.11\pm0.02$ & $0.07\pm0.01$ &
 only $\omega(\rightarrow\pi^0\gamma)\pi^+\pi^-$ based on data \\
$2\pi^+2\pi^-2\pi^0\,({\rm no}\,\eta)$
                    & $1.80\pm0.24$ & $3.32\pm0.29$ &
 purely data based, incl.\ $\eta$ \\
$\pi^+\pi^-4\pi^0\,({\rm no}\,\eta)$
                    & $0.28\pm0.28$ & $0.12\pm0.12$ &
 different relation, incl.\ $\eta$ \\
$\omega(\rightarrow\pi^0\gamma)3\pi$
                    & $0.22\pm0.04$ & -             &
 not estimated \\
$\eta\pi^+\pi^-$ (data) 
                    & $0.98\pm0.24$ & $0.49\pm0.07$ &
 $\eta\rightarrow3\pi$ excluded \\
$\eta\omega$ (data) & $0.42\pm0.07$ & n/a           &
 no data, not estimated separately \\
$\eta\phi$ (data)   & $0.46\pm0.03$ & n/a           &
 no data, not estimated separately \\
$\eta2\pi^+2\pi^-$ (data)
                    & $0.11\pm0.02$ & n/a           &
 no data, not estimated separately \\
$\eta\pi^+\pi^-2\pi^0$
                    & $0.11\pm0.06$ & n/a           &
 not estimated separately \\
\hline
Total               &$12.46\pm0.76$ & $14.25\pm1.46$& \\
\hline
\end{tabular}
\end{center} 
\vspace{-2mm}
\caption{Contributions to $a_\mu$ from exclusive channels for energies
  up $2$ GeV, estimated using isospin relations following
  \cite{Davier:2010nc} and as discussed in the text. For comparison
  also the results of our original analysis are given.} 
\label{tab:iso}
\end{table}

Even after the many improvements due to data from BaBar, there are
still subleading channels where we have only poor or no data at all,
and where the sum of exclusive channels (for energies below $2$ GeV)
requires the use of isospin relations. We have updated the isospin
relations used in our earlier analysis \cite{HMNT03} along the lines
of \cite{Davier:2010nc} and also taking into account the latest data
from BaBar\ \cite{Lees:2011zi}.\footnote{Contrary to the original analyses,
  in the $5\pi$ and $6\pi$ channels the contributions from $\eta$
  decays are excluded, and new channels $\eta\omega$ and $\eta\phi$
  are taken into account, which are now also used in the isospin
  relations. We calculate the contributions from $\eta\omega$ with the
  help of a Breit-Wigner parametrisation, as a data tabulation is not
  available to us.} 
The results for the contributions to $a_\mu$ from the channels
estimated in this way are given in Table~\ref{tab:iso}, where for
comparison also the numbers used for our original analysis
\cite{HMNT03} are shown.\footnote{M3N data in the $\pi^+\pi^-3\pi^0$
  channel, which were used in our original channel compilation, are
  discarded in this analysis, and the $\pi^+\pi^-3\pi^0$ contribution
  is now estimated by isospin relations.} 
Note that the sum of these contributions in
the energy range from $1.43$ to $2$ GeV has gone down by nearly two
units in $10^{-10}$, mainly as a consequence of using new BaBar
data in the $2\pi^+2\pi^-\pi^0$ and the $2\pi^+2\pi^-2\pi^0$ channels as
shown in Figs.\ \ref{fig:ch-14} and \ref{fig:ch-15}.
(The use of the new, more complicated isospin relations
for the multi-pion states has only a minor influence.) Compared to
earlier analyses, recent data from BaBar for the $K\bar K\pi$, $K\bar
K2\pi$ and $K\bar K3\pi$ channels have allowed us to avoid the use of the 
rather badly constrained semi-inclusive channel $K^0_S X$. This has
resulted in a much improved error (by more than a factor of two) of
the contribution from the $K\bar K2\pi$ channel. This channel is
dominating the error in the energy region just below $2$ GeV, which
leads to a significant error reduction as given in the last line of
Table~\ref{tab:iso}.
In future, data from VEPP-2000 in this energy range will provide
important cross-checks for these channels and will hopefully lead to a
further reduction of the error of $a_{\mu}$.

\begin{figure}[htb]
\begin{center}
\psfig{file=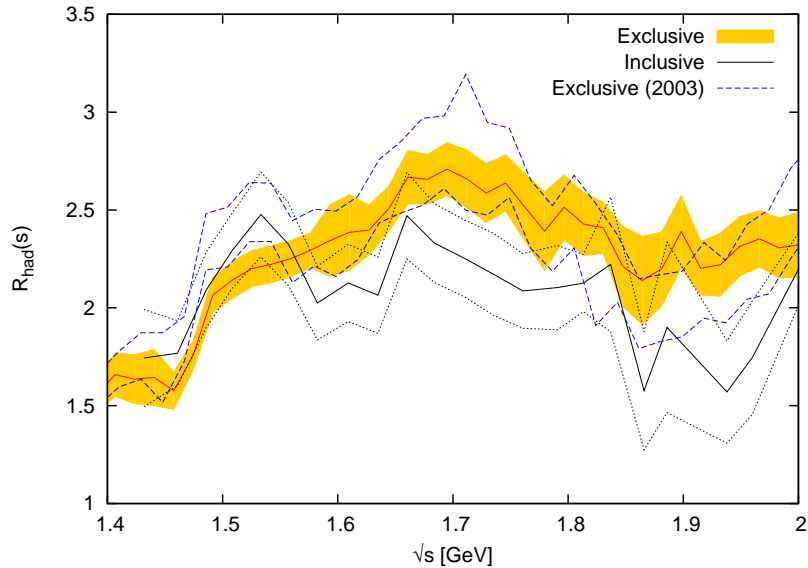,width=11.cm,angle=0}
\end{center}
\vspace{-0.6cm}
\caption{Comparison of our combination of inclusive data with the sum of
  exclusive channels in the energy region below $2$ GeV. The shaded
  band, with the solid (red) line indicating the mean value, gives our
  updated compilation including the new data sets from BaBar; the band
  formed by the two dashed (blue) curves shows our old result used in
  \cite{HMNT03}. The compilation of the old inclusive measurements (solid
  and dotted black lines for mean value and error) has a similar
  energy dependence but is significantly lower over most of the energy
  range.}
\label{fig:inclexcl}
\end{figure}

Note that in our previous $g-2$ analyses \cite{HMNT03,HMNT06} we used two
alternative treatments to calculate the contributions in the $1.43 - 2$
GeV energy interval. We {\em either} added up the contributions from
all the exclusive channels, {\em or} simply used a combination of the
available data on the inclusive cross section, $\sigma(e^+e^- \to
\mbox{all\ hadrons})$. There was a discrepancy between the two methods. The
inclusive data were lower than the sum of the exclusive channels,
though they were found to be similar in shape. Our updated analysis
shows similar findings. This is demonstrated in
Fig.~\ref{fig:inclexcl}, where our compilation for $R_{\rm
  had}(s)=\sigma_{\rm had}^0(s)/(4\pi\alpha^2/(3s))$ from the old
inclusive data is compared to the sum of the exclusive channels as
indicated on the plot and in the caption. The shaded band displays our
new compilation of the sum of the exclusive channels, which is more
accurate, and, for energies up to about $1.8$ GeV, slightly lower than
our older compilation (as used in \cite{HMNT03}) which is shown by the
overlaid dashed (blue) curves. 
In our analysis \cite{HMNT03} the difference in $a_{\mu}$ between
using the `exclusive' and the `inclusive' option was $3.8 \cdot
10^{-10}$, corresponding to about one standard deviation in $a_{\mu}$ in the
energy interval $1.43 < \sqrt{s} <2$ GeV. Quite a few channels were
very poorly constrained by experiment, and for some channels estimates
based on isospin relations had to be used as no data were available
at all. At this point we had invoked a QCD sum-rule analysis, relating
the convolution integral of $R_{\rm had}(s)$ with a suitable kernel
function $f(s)$ to a corresponding contour-integral of the
perturbatively calculated Adler-$D$ function\footnote{Earlier work
  along these lines can be found in \cite{MOR}.}, 
\begin{equation}
\int_{s_{\rm thr}}^{s_0} {\rm d}s\, R_{\rm had}(s)\, f(s) = \int_C {\rm d}s\,
  D(s)\, g(s)\,, 
\label{eq:sumrules}
\end{equation}
with $f(s) = (1-s/s_0)^m (s/s_0)^n$; $C$ is a circular contour of
radius $s_0$ and $g(s)$ is a known function once $f(s)$ is specified
(see \cite{HMNT03} for formulae and details). We then found that the
inclusive data were more compatible with perturbative QCD and the
world-average of $\alpha_s$. Therefore we had preferred to quote the
result obtained from using the inclusive data in this energy region. 

\begin{figure}[htb]
\begin{center}
\psfig{file=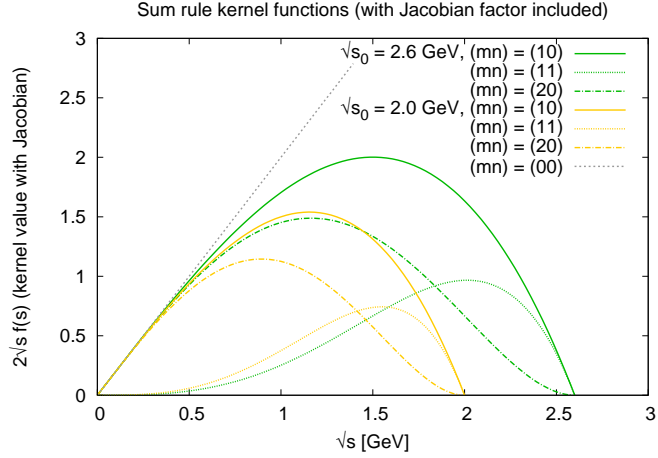,width=9.cm,angle=0}
\end{center}
\vspace{-0.6cm}
\caption{Kernel function $f(s)$, multiplied by the Jacobian factor
  $2\sqrt{s}$, as a function of $\sqrt{s}$ for different choices of the
  parameters $m$, $n$ and $\sqrt{s_0}$, as indicated on the plot.} 
\label{fig:kernels}
\end{figure}
\begin{figure}[htb]
\begin{center}
\psfig{file=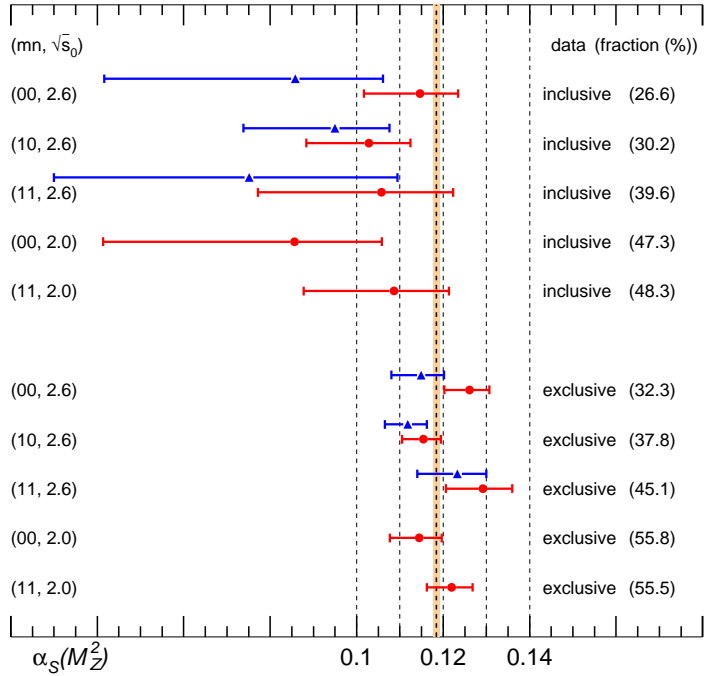,width=11.5cm,angle=0}
\end{center}
\vspace{-1.6cm}
\caption{Sum-rules, (\ref{eq:sumrules}), with different values of $m$,
  $n$ and $\sqrt{s_0}$, translated into a 
  prediction of $\alpha_s$ (see \cite{HMNT03} for details). Upper
  part: use of inclusive data in the region $1.43 < \sqrt{s} < 2$ GeV,
  lower part: use of our up-dated compilation for the sum of exclusive
  data in the same energy interval. 
The (red) circles show the results based on data alone for the left
hand side of (\ref{eq:sumrules}), whereas for the (blue) triangles
perturbative QCD has been used in the region above $2$ GeV, see
subsection~\ref{sec:above2gev}, in particular Fig.~\ref{fig:ch-00}.
The slim (orange) band shows the world-average of
$\alpha_s(M_Z^2)$. Also listed are the fractions with which the energy
region $1.43 < \sqrt{s} < 2$ GeV contributes to the respective
sum-rules.} 
\label{fig:sumrules}
\end{figure}
In light of the significant changes of the data we have now repeated
the sum-rule analysis. For the Adler-$D$ function we now use the full
four-loop result \cite{BaikovCK}, which had not been available for our
original analysis. It is informative to show a range of sum-rules,
although they are highly correlated. The sensitivity of the different
sum-rules w.r.t. the energy range is demonstrated in
Fig.~\ref{fig:kernels}, where different kernel functions $f(s)$
(multiplied by the Jacobian factor) are displayed as a function of the
energy $\sqrt{s}$. The new results are displayed in
Fig.~\ref{fig:sumrules}, where different sum-rules based on the
Adler-$D$ function from pQCD are made to match the corresponding
sum-rule integrals over the data by fitting $\alpha_s$ as a free
parameter. (Note that the error bars for the sum-rule results
represent only the statistical and systematic errors from the
experimental data used in our data compilations. These errors are
large compared to the uncertainty from pQCD when estimated by
comparing the results based on four-loop QCD with those at three-loop
accuracy.) The band displays the current world average for
$\alpha_s(M_Z^2) = 0.1184 \pm 0.0007$ \cite{PDG10}.
\begin{table}[b]
\begin{center}
 \begin{tabular}{l|r|r|c}
 \hline
 \rule{0pt}{2.5ex} 
 Channel & This work & HMNT (03) \cite{HMNT03} &
 Difference \\ 
 \hline
 $\pi^+\pi^-\pi^0\pi^0$ & $ 10.80\pm0.77 $ & $ 10.84\pm0.73 $ & $-0.04$ \\
 $2\pi^+2\pi^-$         & $  8.64\pm0.28 $ & $  8.61\pm0.30 $ & $+0.03$ \\
 $5\pi, 6\pi\,({\rm incl.\,}\eta)$
                        & $  5.92\pm0.41 $ & $  7.65\pm0.43 $ & $-1.73$ \\
 $K\bar{K}\pi$          & $  2.69\pm0.15 $ & $  2.48\pm0.23 $ & $+0.21$ \\
 $K\bar{K}2\pi$         & $  3.31\pm0.58 $ & $  3.63\pm1.32 $ & $-0.32$ \\
 $\pi^+\pi^-\pi^0$      & $  1.25\pm0.07 $ & $  0.61\pm0.09 $ & $+0.64$ \\
 Others                 & $  1.99\pm0.17 $ & $  1.86\pm0.56 $ & $+0.13$ \\
 \hline
 Sum of excl.           & $ 34.61\pm1.11 $ & $ 35.68\pm1.71 $ & $-1.07$ \\
 \hline
 Inclusive          & $ 31.99\pm2.43 $ & $ 31.91\pm2.42 $ & $+0.08$ \\
 \hline
 weighted avg.      & $ 34.15\pm1.10 $ &  &  \\
 \hline
 \end{tabular}
\end{center} 
\vspace{-2mm}
\caption{Contributions to $a_{\mu}$ from the most important channels
  in the region from $1.43$ to $2$ GeV. The numbers given in the second
  column (`This work') are our new results based on the updated
  compilation, whereas the column labelled `HMNT (03)' refers to our old
  analysis \cite{HMNT03}. The last column gives the difference, which,
  due to changes in the treatment of radiative corrections, is also
  present in the combination of the inclusive data, for which no new
  data sets are available. The last three lines give the different
  options for use of data in this region: sum of exclusive channels
  (our preferred choice), inclusive data, or the weighted average.
  (All values in units of $10^{-10}$.)} 
\label{tab:inclexcl}
\end{table}
Compared to our previous analysis in 2003, the sum-rules do not favour
the inclusive data any more, but tend to prefer the more precise
exclusive data. Note that this is partly a consequence of the recent
changes of the low energy (mainly the $\rho$ region) and higher energy
(above $2$ GeV) data common to both options, and partly due to the
change of the data in the range between $1.43$ and $2$ GeV. (The
fraction of the sum-rule stemming from this region is indicated in
brackets in Fig.~\ref{fig:sumrules}.) As a result of these sum-rule
observations, and bearing in mind the recent improvements of the
exclusive data, we are led to use, in the energy range from $1.43$ to
$2$ GeV, the sum over the exclusive data 
for our compilation of $g-2$ and $\Delta\alpha$.\footnote{Also note
  that, in the region just below $2$ GeV, the sum of the exclusive
  channels is in better agreement with the prediction from pQCD than the
  inclusive data. However, we do not use pQCD in this energy region.} 
This does lead to a
considerable (by a factor of two) improvement of the error in this
region but also results in a noticeable shift upwards of the
contributions to $a_{\mu}$ and $\Delta\alpha$ (about $+3\cdot
10^{-10}$ for $a_{\mu}$) compared to our previous analyses. 
(However, note that if we were to average the inclusive and the sum
over exclusive, both the result and the error would change only
marginally w.r.t. using exclusive data only in this region, as
quantified in Table~\ref{tab:inclexcl}.) 

The numerical results are summarised in Table~\ref{tab:inclexcl},
where we display the channels with the largest contributions and the
largest changes in this region. The row labelled `Others' is the sum
of all contributions not listed explicitly.
It is clear that, although the difference between the sum of exclusive
and the inclusive data has not changed dramatically, the new sum of
exclusive channels, with many improvements due to data from BaBar, is
considerably more accurate than the old inclusive data. 

\subsection{Changes in other channels}
\label{sec:otherchannels}
\begin{figure}[htb]
\begin{center}
\psfig{file=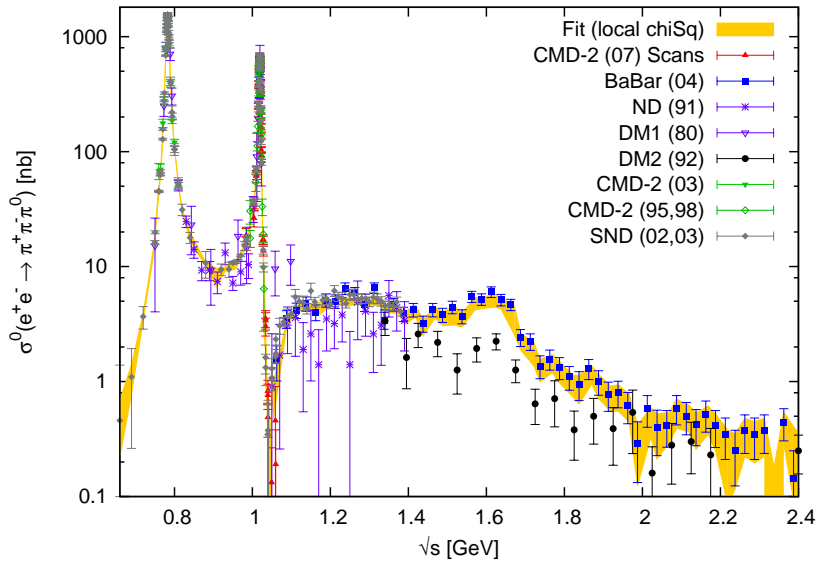,width=11cm,angle=0}
\end{center}
\vspace{-0.6cm}
\caption{$\pi^+\pi^-\pi^0$ channel with $\omega$ and $\phi$ resonances.}
  \label{fig:3pi_all}
\end{figure}
\begin{figure}[htb]
\begin{center}
\psfig{file=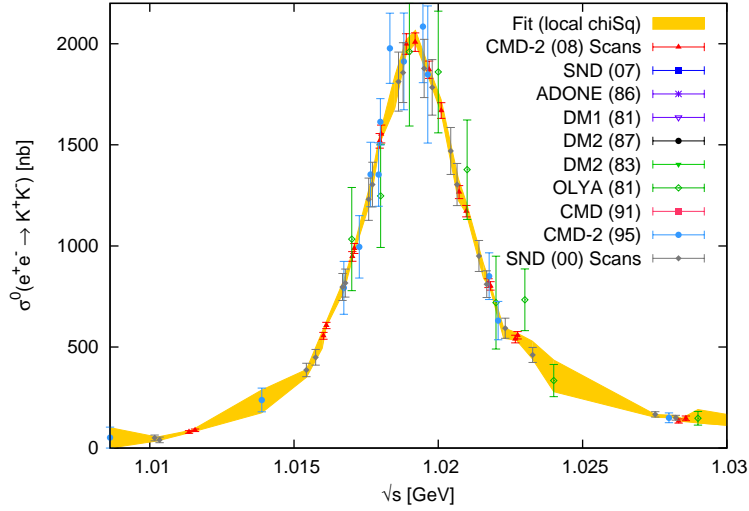,width=10cm,angle=0}
\end{center}
\vspace{-0.6cm}
\caption{$\phi$ resonance in the $K^+K^-$ channel.}
  \label{fig:ch-02_KK}
\end{figure}
In addition to the important improvements in the $2\pi$ channel and in
the region just below $2$ GeV discussed above, there have also been a
number of changes in other channels. Compared to \cite{HMNT06} we have
included the following data sets: $K^+K^-$ from CMD-2 \cite{CMD2_1}
and SND \cite{SND_1}, $K^0_S K^0_L$ from SND \cite{SND_2},
$\pi^+\pi^-\pi^0$ from CMD-2 \cite{CMD2_2}, $\omega\pi^0$ from KLOE
\cite{KLOEomegapi}, and inclusive $R$ data at higher energies above
$2$ GeV from BES \cite{BES_1,BES_2} and CLEO \cite{CLEO}. 
The new data from CMD-2 and SND have further consolidated and improved
the narrow $\omega$ and $\phi$ resonances, whose contributions can be
reliably predicted by direct integration of the data (as opposed to
fits using a specific parametrisation of the resonance). Figure
\ref{fig:3pi_all} shows the $3\pi$ channel with the $\omega$ and
$\phi$ resonances, whereas the $\phi$ resonance in the $K^+K^-$
channel is displayed in Fig.~\ref{fig:ch-02_KK}.

\begin{table}[htb]
\vspace{2mm}
\begin{center}
\begin{tabular}{c|c|c|c}
\hline
Channel & This work & DHMZ (10)\ \cite{Davier:2010nc} & Difference \\
\hline
$\eta\pi^+\pi^-$   & $  0.88\pm0.10$ & $  1.15\pm0.19$ & $-0.27$ \\ 
$K^+K^-$           & $ 22.09\pm0.46$ & $ 21.63\pm0.73$ & $0.46$ \\
$K^0_SK^0_L$       & $ 13.32\pm0.16$ & $ 12.96\pm0.39$ & $ 0.36$ \\ 
$\omega\pi^0$      & $  0.76\pm0.03$ & $  0.89\pm0.07$ & $-0.13$ \\ 
$\pi^+\pi^-$       & $505.65\pm3.09$ & $507.80\pm2.84$ & $-2.15$ \\ 
$2\pi^+2\pi^-$     & $ 13.50\pm0.44$ & $ 13.35\pm0.53$ & $ 0.15$ \\ 
$3\pi^+3\pi^-$     & $  0.11\pm0.01$ & $  0.12\pm0.01$ & $-0.01$ \\ 
$\pi^+\pi^-\pi^0$  & $ 47.38\pm0.99$ & $ 46.00\pm1.48$ & $ 1.38$ \\ 
$\pi^+\pi^-2\pi^0$ & $ 18.62\pm1.15$ & $ 18.01\pm1.24$ & $ 0.61$ \\ 
$\pi^0\gamma$      & $  4.54\pm0.14$ & $  4.42\pm0.19$ & $ 0.12$ \\ 
$\eta\gamma$       & $  0.69\pm0.02$ & $  0.64\pm0.02$ & $ 0.05$ \\ 
$\eta2\pi^+2\pi^-$ & $  0.02\pm0.00$ & $  0.02\pm0.01$ & $ 0.00$ \\ 
$\eta\omega$       & $  0.38\pm0.06$ & $  0.47\pm0.06$ & $-0.09$ \\ 
$\eta\phi$         & $  0.33\pm0.03$ & $  0.36\pm0.03$ & $-0.03$ \\ 
$\phi(\rightarrow{\rm unaccounted})$
                   & $  0.04\pm0.04$ & $  0.05\pm0.00$ & $-0.01$ \\ 
\hline
Sum of isospin channels 
                   & $  5.98\pm0.42$ & $  6.06\pm0.46$ & $-0.08$ \\ 
\hline
Total              & $634.28\pm3.53$ & $633.93\pm3.61$ & $ 0.35$ \\ 
\hline
\end{tabular}
\end{center} 
\vspace{-2mm}
\caption{Contributions to $a_\mu$ (in units of $10^{-10}$) in the
  energy region from $0.305$ to $1.8$ GeV from exclusive channels:
  Results based on the data compilation as used in this analysis
  compared to the results as given by Davier {\it et al.}\ 
  \cite{Davier:2010nc}.} 
\label{tab:comparisonDavieretal}
\end{table}
In Table~\ref{tab:comparisonDavieretal} we compare our predictions of
the contributions to $a_\mu$ from the exclusive channels up to $1.8$
GeV with the results as given in \cite{Davier:2010nc}. While the
agreement of the sum of all contributions is very good, there are 
differences in several channels, which are of the order of the errors,
notably in the $2\pi$, $3\pi$ and the $KK$ channels. These differences
presumably come from a different treatment of the data, like radiative
corrections and the clustering and integration procedure, and reflect
the different choices made by different groups.

\begin{figure}[htb]
\begin{center}
\psfig{file=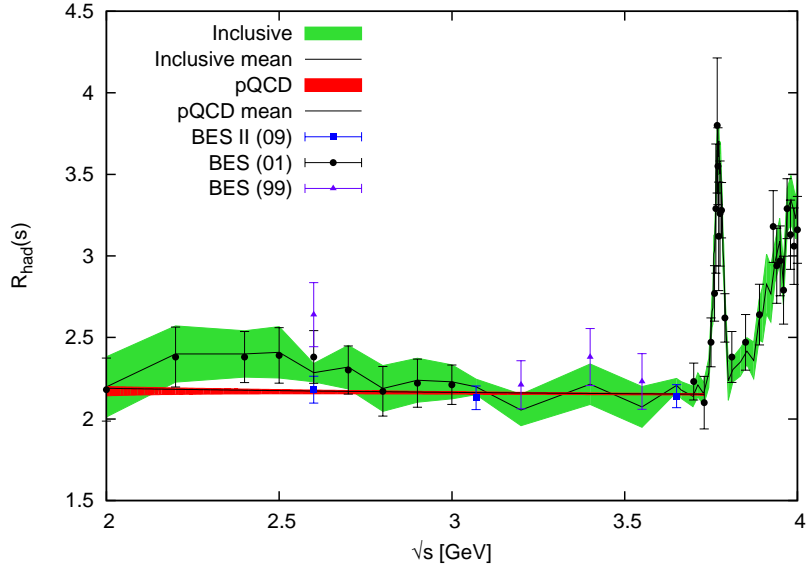,width=11.cm,angle=0}
\end{center}
\vspace{-0.6cm}
\caption{Compilation of inclusive data (light (green) band) above $2$
  GeV compared to the prediction of pQCD (dark (red) band). The peaked
  structure is the $\psi(3770)$, very close to the open charm threshold. The
  $J/\psi$ and $\psi^{\prime}$ are not resolved by these data but are
  included separately as narrow resonances in our analysis.} 
  \label{fig:ch-00}
\end{figure}
\subsection{Region above $2$ GeV}
\label{sec:above2gev}
Of particular interest is the region of inclusive data above $2$
GeV. Although its influence for $g-2$ is suppressed by the kernel
function $K$ of (\ref{eq:disp}), the new data lead to a significant
shift downwards of the hadronic contributions to $g-2$ from this
energy region. (The direct integration of our new data compilation in
the energy range $2 < \sqrt{s} < 11.09$ GeV gives an $a_{\mu}$
contribution lower by about $1.4 \cdot 10^{-10}$ compared to our
earlier result in \cite{HMNT06}). 
In Fig.~\ref{fig:ch-00} the new fit and some of the important data are
compared to pQCD. It is intriguing that the newest data from BES
\cite{BES_2}, which are responsible for the pull-down in this
region, are in perfect agreement with the predictions from pQCD, which
in turn are considerably lower than the data in the region around $2.4$
GeV. 
It is interesting to 
replace data by pQCD above $2$ GeV in the sum-rule analysis. We have
done so by employing the latest version of the routine {\tt rhad}
\cite{Harlander:2002ur} which includes recently obtained four-loop
contributions, and by leaving $\alpha_s$ as a free parameter in the
sum-rule analysis. Our results are displayed in
Fig.~\ref{fig:sumrules} as (blue) triangles, which are, as
expected, lower than the respective results based on data
alone.
The sum-rule integrals, expressed in terms of $\alpha_s(M_Z^2)$, now
show a slightly improved agreement with the world-average for the
exclusive data, and an even worse agreement if inclusive data are
used. Nevertheless, for our data-driven analysis of $g-2$ and
$\Delta\alpha$, we use the data in the region from $2$ to $2.6$
GeV. However, between $2.6$ and $3.73$ GeV we do use pQCD, but with an
inflated error determined by the percentage errors of the BES data
\cite{BES_2}.\footnote{If we would use pQCD also from $2 < \sqrt{s} <
  2.6$ GeV and estimate the error from the pQCD uncertainty, the
  contribution to $a_{\mu}$ from this region would be $(14.49 \pm
  0.13) \cdot 10^{-10}$ instead of $(15.69 \pm 0.63) \cdot
  10^{-10}$. If we would use (inclusive) data instead of pQCD from
  $2.6 < \sqrt{s} < 3.73$ GeV, our prediction for $a_\mu$ would be
  larger by about $0.2 \cdot 10^{-10}$ with a marginally increased
  error. Using pQCD instead of data at higher energies above
  the charm resonance region but below the bottom threshold would lead
  to a negligible change in $a_\mu$.} 
From $11.09$ GeV we use pQCD as done already in \cite{HMNT03}. The
data in this region is very poor and pQCD, above the open bottom
threshold, is very well justified. Note that, as in
\cite{HMNT03,HMNT06}, the narrow resonances $J/\psi$, $\psi^{\prime}$
and the $\Upsilon(1-6 S)$ are added separately as they are not
resolved by the data used in the inclusive compilation. 

\subsection{$a_{\mu}^{\rm had,\,LO\,VP}$ and $a_{\mu}^{\rm had,\,HO\,VP}$}
\label{sec:vpfull}
Table~\ref{tab:amuvplo} gives the contributions to the leading order
hadronic vacuum polarisation from different energy ranges, including
different options for the region from $1.43$ to $2$ GeV, and makes the
comparison with our earlier results from \cite{HMNT06}.
\begin{table}[htb]
\begin{center}
 \begin{tabular}{l|r|r|c}
 \hline
 \rule{0pt}{2.5ex}
 Contribution & This work & HMNT (06) \cite{HMNT06} &
 Difference \\
 \hline \hline
 $2m_{\pi}-0.32\,{\rm GeV}\,({\rm ChPT},\,2\pi)$ 
                                           & $   2.36\pm0.05 $ & $
                                           2.36\pm0.05 $ & $\pm 0.00$ \\ 
 $3m_{\pi}-0.66\,{\rm GeV}\,({\rm ChPT},\,3\pi)$
                                           & $   0.01\pm0.00 $ & $
                                           0.01\pm0.00 $ & $\pm 0.00$ \\ 
 $m_{\pi}-0.60\,{\rm GeV}\,({\rm ChPT},\,\pi^0\gamma)$
                                           & $   0.13\pm0.01 $ & $
                                           0.13\pm0.01 $ & $\pm 0.00$ \\ 
 $m_{\eta}-0.69\,{\rm GeV}\,({\rm ChPT},\,\eta\gamma)$
                                           & $   0.00\pm0.00 $ & $
                                           0.00\pm0.00 $ & $\pm 0.00$ \\ 
 $\phi\rightarrow\,$unaccounted modes           & $   0.04\pm0.04 $ &
 $   0.06\pm0.06 $ & $- 0.02$ \\ 
 $0.32-1.43\,{\rm GeV}$                    & $ 606.50\pm3.35 $ & $
 601.96\pm3.19 $ & $+4.54$ \\ \hline
 $1.43-2\,{\rm GeV}\,$(excl.\ only)       & $  34.61\pm1.11 $ & $
 36.38\pm1.66 $ & $-1.77$ \\ 
 $1.43-2\,{\rm GeV}\,$(incl.\ only)       & $  31.99\pm2.43 $ & $
 32.05\pm2.43 $ & $-0.06$ \\ 
 $1.43-2\,{\rm GeV}\,$(incl.-excl.\ avg.) & $  34.15\pm1.10 $ & n/a &
 n/a \\ \hline
 $2-11.09\,{\rm GeV}$                   & $  41.19\pm0.82 $ & $
 42.75\pm1.08 $ & $-1.56$ \\ 
 $J/\psi+\psi'$                            & $   7.80\pm0.16 $ & $
 7.90\pm0.16 $ & $- 0.10$ \\ 
 $\Upsilon(1{\rm S}-6{\rm S})$             & $   0.10\pm0.00 $ & $
 0.10\pm0.00 $ & $\pm 0.00$ \\ 
 $11.09-\infty\,$(pQCD)                    & $   2.11\pm0.00 $ & $
 2.11\pm0.00 $ & $\pm 0.00$ \\ 
 \hline \hline
 Sum (excl.--excl.--incl.)                 & $ 694.86\pm3.71 $ & $ 693.77\pm3.84 $ & $+1.09$ \\
 Sum (excl.--incl.--incl.)                 & $ 692.25\pm4.23 $ & $ 689.44\pm4.17 $ & $+2.81$ \\
 Sum (excl.--avg.--incl.)                  & $ 694.40\pm3.67 $ & n/a & n/a \\
 \hline
 \end{tabular}
\end{center}
\vspace{-2mm}
\caption{
Contributions to $a_{\mu}^{\rm had,\,LO\,VP}$ obtained in
  this work compared to the values used in our analysis
  \cite{HMNT06}. The last column gives the differences. (All values in
  units of $10^{-10}$.) The first four lines give our predictions of
  contributions close to threshold where no data are available and are
  based on chiral perturbation theory (ChPT), see \cite{HMNT03} for
  details. For $2.6 < \sqrt{s} < 3.73$ GeV pQCD with errors comparable
  to those of the latest BES data is used as default for this work,
  see the discussion in the text. The different choices quoted in the
  last three lines refer to the energy regions below $1.43$ GeV, for
  $1.43 < \sqrt{s} < 2$ GeV and above.} 
\label{tab:amuvplo}
\end{table}
From the fourth column it is clear that the changes w.r.t.\ our
analysis from 2006 partially cancel each other. However, the inclusion
of the radiative return data from KLOE and BaBar, and taking into
account the exclusive data in the region from $1.43$ to $2$ GeV,
dominates the changes and leads to a slightly increased prediction of
$a_{\mu}^{\rm had,\,LO\,VP}$ of
\begin{equation}
a_{\mu}^{\rm had,\,LO\,VP} = (694.91 \pm 3.72_{\rm exp} \pm 2.10_{\rm
  rad}) \cdot 10^{-10} \,.
\label{eq:amuvplo}
\end{equation}
The first error, labelled `exp', stems from the statistical and
systematic errors of the experimental data, as used in our
combination procedure. The second, additional error, labelled `rad', is due to 
uncertainties in the application of radiative corrections to the
data. For a detailed discussion of its estimate see \cite{HMNT03}. 
Note that the value given in (\ref{eq:amuvplo}) slightly differs from
the one quoted in Table~\ref{tab:amuvplo}, for which the $2\pi$ data were
used from $0.32$ GeV to facilitate the comparison with
\cite{HMNT06}. For our new prediction (\ref{eq:amuvplo}), the $2\pi$
data are used from $0.305$ GeV and only below this chiral perturbation
theory is applied. 

With the same data compilation we can also determine the higher order
VP contributions, see \cite{HMNT03} for details concerning the
corresponding dispersion integrals. Our new value is only slightly
changed from our previous prediction and reads
\begin{equation}
a_{\mu}^{\rm had,\,HO\,VP} = (-9.84 \pm 0.06_{\rm exp} \pm 0.04_{\rm
  rad}) \cdot 10^{-10} \,.
\label{eq:amuvpho}
\end{equation}
Equations~(\ref{eq:amuvplo}) and (\ref{eq:amuvpho}) are used for our
updated prediction of $g-2$. These results, together with
(\ref{eq:delal}), are the main results of this paper. 

\section{Standard Model Prediction of $g-2$}
\label{sec:gm2comp}
For the Standard Model prediction of $(g-2)_{\mu}$, contributions from all
sectors have to be added:
\begin{equation}
a_{\mu}^{\rm SM} = a_{\mu}^{\rm QED} + a_{\mu}^{\rm EW} + a_{\mu}^{\rm had}\,.
\label{eq:smsectors}
\end{equation}
In contrast to the hadronic sector, both QED and electro-weak (EW)
contributions can be calculated reliably using perturbation theory. 
After many years' work the QED contributions are known to full
four-loop accuracy, and estimates for the five-loop
contributions are ongoing (see e.g.\ the recent works
\cite{Aoyama:2008hz,Aoyama:2010yt,Aoyama_latest}). Below we will use the value 
$a_{\mu}^{\rm QED} = 116584718.08(15) \cdot 10^{-11}$
\cite{Kinoshitaetal,Aoyamaetal}, where the error is dominated by the
estimate of the unknown five-loop contributions (for a detailed
discussion and more references see e.g.\ the recent review
\cite{Jegerlehner:2009ry}). 
The EW corrections are known to two-loop accuracy
\cite{Czarnecki:2002nt,Czarnecki:1995sz,Czarnecki:1995wq,Knecht:2002hr,Peris:1995bb}
and amount to  $a_{\mu}^{\rm EW} = (154 \pm 2) \cdot 10^{-11}$, where
the error estimate is due to the remaining hadronic uncertainties, the
unknown Higgs mass and undetermined higher-order
contributions. Clearly, compared with the uncertainties of the VP
contributions discussed above, both the QED and EW corrections are
very well under control. In the hadronic sector, as well as the
(LO and HO) VP corrections, we also have to take into account the
light-by-light scattering contributions. They enter at the
same order $\alpha^3$ as the HO VP corrections, but can not be
determined from data via dispersive methods. 
All model-based estimates include the pseudoscalar contributions,
i.e.\ exchanges of $\pi^0$, $\eta$ and $\eta^{\prime}$, which are
leading in the large $N_c$ limit. In addition, axial vector exchanges,
charged $\pi$ and $K$ loops, and (dressed) quark loop diagrams are
taken into account, and short-distance constraints from pQCD have been
applied to enforce a consistent matching at higher virtualities. 
Although there are important differences in the treatment of the
different contributions, the recent results $a_{\mu}^{\rm had,\,l-by-l} =
(10.5 \pm 2.6) \cdot 10^{-10}$ \cite{Prades:2009tw} and $a_{\mu}^{\rm had,\,
  l-by-l} = (11.6 \pm 4.0) \cdot 10^{-10}$
\cite{Nyffeler:2009tw,Jegerlehner:2009ry} turn out to be compatible
(see also \cite{Nyffeler:2010rd} for a recent short review). 
Note that the recent results from \cite{Prades:2009tw} and
\cite{Nyffeler:2009tw} agree fairly well w.r.t.\ the leading
contributions, and that both have cancellations in the subleading
parts, thus strengthening our confidence in the reliability of these
estimates. Also note that these predictions of $a_{\mu}^{\rm had,\,l-by-l}$
are below the estimated upper bound $a_{\mu}^{\rm had,\,l-by-l} < 15.9
\cdot 10^{-10}$ based on parton hadron duality
\cite{Erler:2006vu}. For our prediction of $a_{\mu}^{\rm SM}$ we will
use the result from \cite{Prades:2009tw}, which has been obtained as a
`best estimate' for $a_{\mu}^{\rm had,\,l-by-l}$ after reviewing different
approaches. In the future it may well be possible to obtain
independent constraints on, or hopefully even a full prediction of, the
light-by-light contributions from lattice gauge field theory. Such
first principles simulations of the required four-current correlator
are very difficult, but work by two groups is underway
\cite{Blum:2009zz,PRakow} and the first steps are
encouraging.\footnote{For first results obtained within an alternative
  approach based on Dyson-Schwinger methods see Goecke {\it et
    al.}~\cite{Goecke:2010if}. They estimate $a_{\mu}^{\rm
    had,\,l-by-l}$ to be largely enhanced by quark loop
  contributions. However, see \cite{Boughezal:2011vw} for a
  counter-argument.} 
In addition, measurements of the meson form factors, which are
needed in the modelling of the light-by-light contributions, may become
feasible at several experiments at low energy $e^+e^-$ colliders.

\begin{figure}[htb]
\begin{center}
\psfig{file=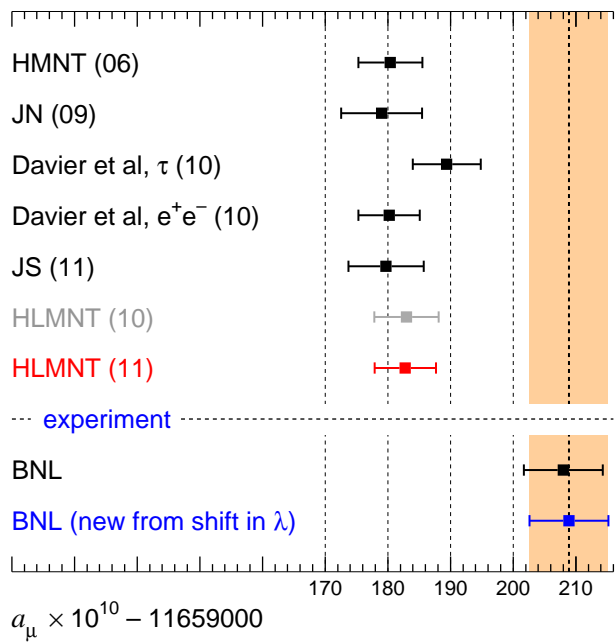,width=10cm,angle=0}
\end{center}
\vspace{-1.cm}
\caption{Standard model predictions of $a_\mu$ by several groups
  compared to the measurement from BNL
  \cite{BNL,Roberts:2010cj,Amsler:2008zzb}. The SM predictions are
  from HMNT (06) \cite{HMNT06}, JN (09) \cite{Jegerlehner:2009ry},
  Davier {\it et al.} \cite{Davier:2010nc}, JS (11) \cite{JS}, HLMNT
  (10) \cite{HLMNTprocs2010}, and HLMNT (11) is this work. Note that
  the value from Jegerlehner and Szafron includes $\tau$ spectral
  function data, which, in their approach, are fully consistent with
  and confirm the $e^+e^-$ data. HLMNT (10) is a preliminary version
  of this work, presented at conferences \cite{HLMNTprocs2010}, but
  before the full updated data set was available.} 
\label{fig:asm}
\end{figure}
Adding all the hadronic, QED and EW contributions, we finally arrive
at the SM prediction
\begin{equation}
a_{\mu}^{\rm SM} =  (11\,659\,182.8 \pm 4.9) \cdot 10^{-10} \,,
\label{eq:amusm}
\end{equation}
where the errors have been added in quadrature. This prediction is now
even slightly more precise than the seminal experimental measurement
from BNL \cite{BNL}. After taking into account the small shift
in CODATA's published ratio of the muon-to-proton magnetic moment
\cite{Mohr:2008fa}, the experimental value for $a_{\mu}$ reads
\cite{Roberts:2010cj,Amsler:2008zzb} 
\begin{equation}
a_{\mu}^{\rm EXP} = 116\,592\,089(63) \cdot 10^{-11}\,.
\end{equation}
This leads to the difference
\begin{equation}
a_{\mu}^{\rm EXP} - a_{\mu}^{\rm SM} =  (26.1 \pm 8.0) \cdot 10^{-10}
\label{eq:amudiff}
\end{equation}
which corresponds to $3.3\,\sigma$ discrepancy. In Fig.~\ref{fig:asm}
we compare the SM predictions of different groups (markers as
indicated) and the BNL measurement, which is displayed by the (orange)
band. Note that, despite many changes in the recent history, the
different predictions agree very well with each other, and the
discrepancy seen in this observable is persisting. While, at a level
of $3\,\sigma$, one can not speak of a firmly established deviation
from the SM prediction, all different contributions have been checked
thoroughly. It should also be noted that it seems increasingly
difficult to explain the discrepancy by a change in the hadronic data
alone, as this would lead to increased tension with the EW precision
fits of the SM and the current Higgs mass limits
\cite{Passera:2008jk,Passera:2010ev}. 

\section{$\Delta\alpha(M_Z^2)$}
\label{sec:delal}
Leptonic and hadronic vacuum polarisation effects screen the electric
charge and lead to the scale dependence (`running') of the QED coupling,
$\alpha(q^2) = \alpha/(1-\Delta\alpha_{\rm lep}(q^2)-\Delta\alpha_{\rm
  had}(q^2))$, where $\alpha$ is the fine structure constant and
$\Delta\alpha(q^2)$ the real part of the vacuum polarisation of the
photon. Of particular importance is the value at the scale of the mass
of the $Z$ boson, $\alpha(M_Z^2)$. Remarkably, as a consequence of the
hadronic uncertainties, this is the least well known of the EW SM
parameters, $\left[G_{\mu}, M_Z, \alpha(M_Z^2)\right]$. It is needed
for precise predictions of high-energy processes and is a crucial
ingredient in the EW precision fits of the SM, which in turn lead to
the indirect determination of the Higgs mass. 

We can use the same compilation of hadronic data (combined with
pQCD) to calculate, with the best possible precision, the hadronic
contributions to the running of the QED coupling from the dispersion
integral 
\begin{equation}
\Delta\alpha_{\rm had}(q^2) = -\frac{\alpha q^2}{3\pi}\, P \int_{s_{\rm
    th}}^{\infty} \frac{R_{\rm had}(s^{\prime}){\rm
      d}s^{\prime}}{s^{\prime}(s^{\prime}-q^2)}\,,
\label{eq:delalint}
\end{equation}
where $P$ denotes the principal value of the integral. Above $11.09$
GeV we use pQCD 
to calculate $R_{\rm had}$. As usual, the
contribution from the top quark is added separately. Our new result
for the five-flavour hadronic contributions then reads 
\begin{equation}
\Delta\alpha_{\rm had}^{(5)}(M_Z^2) = \left( 276.26 \pm 1.16_{\rm exp}
  \pm 0.74_{\rm rad} \right) \cdot 10^{-4}\,. 
\label{eq:delal}
\end{equation}
Together with the well known analytic results for the leptonic and the
top quark contributions, $\Delta\alpha_{\rm lep}(M_Z^2) = 0.031498$
\cite{Steinhauser:1998rq} and
$\Delta\alpha_{\rm top}(M_Z^2) = -0.0000728 (14)$ \cite{Delaltop}
(where we have used $m_t = (172.0 \pm 1.6)$ GeV \cite{PDG10}),
this allows for the further improved prediction
\begin{equation}
\alpha(M_Z^2)^{-1} = 128.944 \pm 0.019\,.
\label{eq:aqedmz}
\end{equation}
\begin{table}[htb]
\begin{center}
 \begin{tabular}{l|l|l}
 \hline
 \rule{0pt}{2.5ex}
 Group, year, ref. & $\Delta\alpha_{\rm had}^{(5)}(M_Z^2)$ & Remarks \\ 
 \hline\hline
K\"uhn+Steinhauser (98) \cite{Kuhn:1998ze} & $0.02775 \pm 0.00017$ &
pQCD \\ \hline 
Martin {\it et al.} (00) \cite{MOR} & $0.02738 \pm 0.00020$ & data
driven \\ \hline
 Troconiz+Yndurain (05) \cite{de Troconiz:2004tr} & $0.02749 \pm
 0.00012$ & pQCD \\ \hline 
 Burkhardt+Pietrzyk (05) \cite{Burkhardt:2005se} & $0.02758 \pm
 0.00035$ & data driven \\ \hline
HMNT (06) \cite{HMNT06} & $0.02768 \pm 0.00022$ & data driven\\ \hline
Jegerlehner (08) \cite{Jegerlehner:2008rs} & $0.027594 \pm 0.000219$ &
data driven/pQCD\\ 
 & $0.027515 \pm 0.000149$ & Adler function ($\sqrt{s_0} = 2.5\ {\rm GeV}$)\\  
Jegerlehner (10) \cite{FJ2010} & $0.027498 \pm 0.000135$ & Adler
function ($\sqrt{s_0} = 2.5\ {\rm GeV}$)\\ \hline
Davier {\it et al.} (10) \cite{Davier:2010nc} & $0.02742 \pm 0.00010$ &
pQCD from $1.8 < \sqrt{s} < 3.7$ GeV\\
 & & and for $\sqrt{s} > 5$ GeV\\ \hline
HLMNT (11), this work & $ 0.027626 \pm 0.000138 $ & data driven\\
 \hline
 \end{tabular}
\end{center} 
\vspace{-2mm}
\caption{Results for $\Delta\alpha_{\rm had}^{(5)}(M_Z^2)$ from
  different groups. The column `remarks' indicates if the analysis is
  mainly relying on data as input in the dispersion integral
  (\ref{eq:delalint}) or if pQCD is used outside the resonance
  regions; another approach proposed by Jegerlehner is based on the
  use of the Adler $D$ function, thus reducing the dependence on data
  and improving the error.} 
\label{tab:delalmz}
\end{table}
\begin{figure}[htb]
\begin{center}
\psfig{file=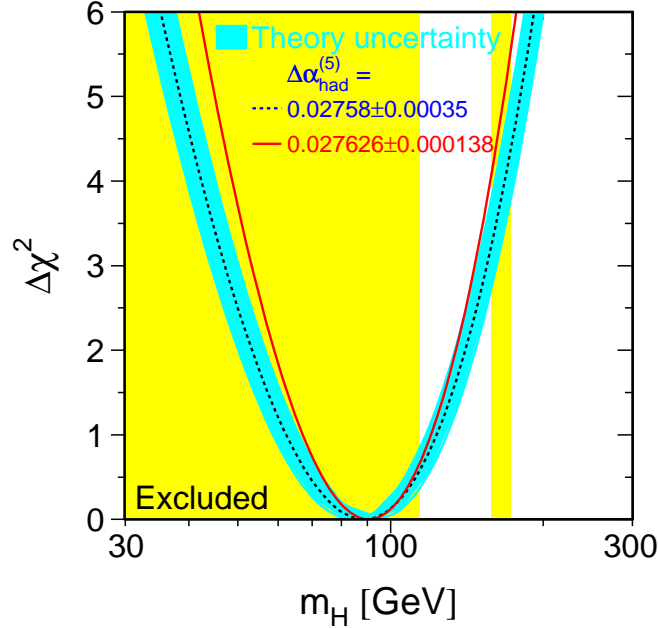,width=9cm,angle=0}
\end{center}
\vspace{-1.2cm}
\caption{`Blue-band plot' from the LEP EWWG: The dark (blue) band with
  the dashed line shows the default EW precision data fit result (July
  2010), whereas the solid (red) parabola is obtained by using our new
  value for $\Delta\alpha_{\rm had}^{(5)}(M_Z^2)$ \cite{MGrunewald}.} 
\label{fig:blueband}
\end{figure}
\vspace{-5mm}
In Table~\ref{tab:delalmz} we compare our result for $\Delta\alpha_{\rm
  had}^{(5)}(M_Z^2)$ with recent evaluations from other groups. Note
that our result is slightly bigger and significantly more accurate than the
prediction \cite{Burkhardt:2005se} used as default by the LEP
Electroweak Working Group for their precision fits \cite{EWWG}. In the
so-called `blue-band plot', which summarises the indirect determination
of the Higgs mass, using our new value for $\alpha(M_Z^2)$ results in
a steeper parabola, see the solid (red) line in
Fig.~\ref{fig:blueband}. The fit result for the Higgs mass is $m_H =
\left( 91\ ^{+30}_{-23} \right)$ GeV \cite{MGrunewald}, which is to be
compared to $m_H = \left(89\ ^{+35}_{-26} \right)$ GeV for the default
fit (July 2010). Together with the regions excluded by direct Higgs
searches at LEP and Tevatron (indicated by the shaded (yellow) areas in
Fig.~\ref{fig:blueband}), this could hint at another problem for the
SM, similar to, but less significant than the discrepancy in $g-2$. 

We have also written an easy-to-use Fortran package\footnote{The whole
  package is available upon request from the authors as a
  self-contained collection of Fortran routines and includes the real and
  imaginary part of the leptonic and hadronic vacuum polarisation.} for
the running coupling at space- and time-like momenta, which will be
discussed in a separate work \cite{HLMNT2} (see also chapter 6 in
\cite{MCSIGHAD} for a recent review). Note that for the analyses of
$g-2$ and $\Delta\alpha$ presented here we have used, for consistency,
our own routine for corrections of the data w.r.t. VP effects. 

\section{Conclusions and Outlook}
\label{sec:sum}

\begin{figure}[htb]
\begin{center}
\psfig{file=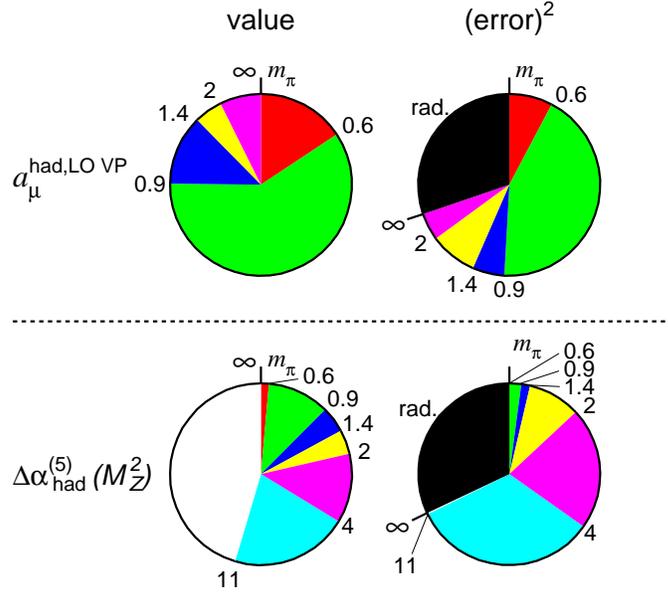,width=11cm,angle=0}
\end{center}
\vspace{-2.4cm}
\caption{The pie diagrams in the left- and right-hand columns show 
  the fractions of the total contributions and $({\rm errors})^2$,
  respectively, coming from various energy intervals in the dispersion 
  integrals (\ref{eq:disp}) and (\ref{eq:delalint}). The pie 
  diagrams for the LO hadronic contribution to $g-2$, shown in 
  the first row, correspond to sub-contributions with energy boundaries at
  $m_\pi, 0.6, 0.9, 1.4, 2~\GeV {\rm ~and}~\infty$, whereas for the
   hadronic contribution to the QED coupling, shown in the second row, 
  the boundaries are at 
  $m_\pi, 0.6, 0.9, 1.4, 2, 4, 11.09~\GeV {\rm ~and}~\infty$.
  In the $({\rm error})^2$ pie diagrams we also included the
  $({\rm error})^2$ arising from the treatment of the radiative corrections
  to the data.}
\label{fig:pie}
\end{figure}

Here we have been concerned with obtaining the best possible accuracy
for the theoretical predictions for the anomalous magnetic moment of
the muon, $(g-2)/2$, and for the QED coupling at the $Z$-boson mass,
$\alpha(M_Z^2)$; two quantities which play a very important role in
electroweak precision physics. Indeed, the discrepancy between the SM
prediction and the experimental measurement of $g-2$ is arguably the
best hint, at present, of physics beyond the SM. At the moment, the
accuracy of the predictions are limited by the uncertainty of the
hadronic vacuum-polarisation contributions. We have used all the
available data on $e^+e^- \to \mbox{hadrons}$ to achieve a data-driven
determination of these contributions, as accurately as possible. We
employ a detailed cluster algorithm, with local $\chi^2$ inflation
when necessary, to combine data from different experiments in a
reliable and consistent way. From the pie diagrams in the first row of
Fig.~\ref{fig:pie} we see that the dominant contribution to the LO
hadronic vacuum-polarisation correction, $a_{\mu}^{\rm had,LO~VP}$, to
$g-2$  comes from the $e^+e^-$ energy region up to 0.9 GeV, whereas
the major part of the error comes from the region up to $2$ GeV. 

Recently new data for many exclusive channels in the low to
intermediate energy region have become available, both from
direct scan experiments and from analyses of radiative return
measurements. In the $2\pi$ channel, the different experiments show
slight disagreement. In particular, there is tension between the
radiative return results obtained from the KLOE and BaBar experiments,
with the BaBar data significantly higher than the KLOE data especially
at larger energies. The tension prevents a bigger improvement of the
error on the contribution from this important region than might have
been expected, given the increased statistics; but it also means that
there clearly is still scope for improved accuracy from future
measurements. 

There is the possibility of evaluating the contribution in the
$e^+e^-$ energy region $1.43-2$ GeV, {\it either} by using the data
for the inclusive $e^+e^- \to \mbox{hadrons}$ cross section, {\it or} by
using the sum of the cross sections for all the exclusive $e^+e^-$
cross sections. There is a discrepancy between the two alternatives,
and, in the past, we had used a sum-rule analysis \cite{HMNT03} to
distinguish between them. However, in the region below $2$ GeV, many
exclusive channels with higher multiplicities are now much better
probed through the radiative return analyses of the BaBar data, and
the sum of the exclusive channels is better determined. This has
prompted us to re-investigate the sum-rule analysis including the new
data in the input. We now find good agreement of the sum-rules with
the world average value of $\alpha_s$ if the sum of exclusive channels
is used, which is slightly higher than the inclusive data, and a
worsened agreement if inclusive data are input. We therefore now use
the exclusive data, which are also more accurate. 

Furthermore, new much more accurate BES data are seen to be in perfect
agreement with the pQCD predictions of $e^+e^- \to \mbox{hadrons}$ in the
range from $2.6$ GeV up to the charm threshold. We therefore use pQCD in
this region, but with a conservative error of about $3.5\%$
corresponding to the accuracy of the latest BES data. Use of pQCD from
$2$ GeV would result in a slight shift ($-1.2\cdot 10^{-10}$ for
$a_\mu$), with an even stronger preference for the exclusive data in
the sum-rule analysis, see Fig.~\ref{fig:sumrules}. 

In summary, we find the updated LO and HO hadronic vacuum-polarisation
corrections to be 
\begin{equation}
a_{\mu}^{\rm had,\,LO\,VP} = (694.91 \pm 4.27) \cdot 10^{-10} \,,
\end{equation}
\begin{equation}
a_{\mu}^{\rm had,\,HO\,VP} = (-9.84 \pm 0.07) \cdot 10^{-10} \,.
\end{equation}
When the representative value of the hadronic light-by-light correction,
\begin{equation}
a_\mu^{\rm had, l-by-l} = (10.5 \pm 2.6) \cdot 10^{-10}\,,
\end{equation}
and the QED and EW contributions are added, our SM prediction for the
anomalous magnetic moment of the muon is 
\be 
a_{\mu}^{\rm SM} =  (11\,659\,182.8 \pm 4.9) \cdot 10^{-10} \,,
\ee
which is a slight shift and improvement of our earlier result \cite{HMNT06},
$a_{\mu}^{\rm SM}({\rm HMNT\ 06}) =  (11\,659\,180.4 \pm 5.1) \cdot
10^{-10} \,$. 
More important, despite many changes in the data, the SM prediction of
$g-2$ 
is further consolidated, and with it the discrepancy of $3.3\,\sigma$
between the BNL measurement and the SM prediction. Moreover, there is
also agreement with other independent analyses, which use somewhat
different treatments of the data. 

There is great potential for further improvements in both the
experimental and theoretical determinations of $g-2$ of the muon. On
the experimental side, the experiments CMD-3 and SND at VEPP-2000 in
Novosibirsk have already started taking data. More radiative return
analyses from BaBar and KLOE are in preparation. Also more
measurements from BES are anticipated. In the longer term, there are
prospects for an upgrade of DA$\Phi$NE and for Super$B$ factories. These
experiments will allow a further reduction of the error on the
calculation of the hadronic vacuum polarisation contributions, which
will be essential in view of the next generation measurements of $g-2$
planned at Fermilab and at J-PARC. In fact, then the error on the
calculation of the hadronic light-by-light contributions will become
critical. Indeed, for the theoretical prediction to continue to match
the experimental measurement of $g-2$ in precision, it is clear that
theoretical efforts, together with measurements of the two-photon form
factors of scalars and pseudo-scalars, are needed to improve the model
predictions in the light-by-light sector. 
The $g-2$ discrepancy has implications for constraining SUSY
models as we discussed in \cite{HMNT03}; for a very recent application
and further references see \cite{ChoHagiwaraMatsumotoNomura}.

As a by-product of the $g-2$ analysis, we used the same compilation of
hadronic data to update the prediction of the value of the QED
coupling, whose value at the $Z$ scale is
\be
\alpha(M_Z^2)^{-1} = 128.944 \pm 0.019\,.
\ee
In this case, we see from Fig.~\ref{fig:pie} that the contributions to
$\,\Delta\alpha_{\rm had}^{(5)}(M_Z^2) = \left( 276.26 \pm 1.38
\right) \cdot 10^{-4}\,$ have quite a different dependence on the
$e^+e^-$ energy, with much less reliance on the lower energies. When
this value for $\Delta\alpha_{\rm had}^{(5)}(M_Z^2)$ is used in the EW
precision fits of the SM, the Higgs mass is constrained more tightly,
with its optimum value in a region excluded by the direct searches,
namely $m_H = \left(91\ ^{+30}_{-23} \right)$ GeV at 68\% confidence
level. 

\vspace{1cm}
\noindent
{\bf\large Acknowledgments}

\vspace{4mm}
\noindent
We thank Martin Gr\"unewald for providing us with the `blue-band plot'
using our value of $\Delta\alpha_{\rm had}^{(5)}(M_Z^2)$. We also
thank M. Davier, S. Eidelman, A. Hoecker, F. Jegerlehner, B. Malaescu
and G. Venanzoni for valuable discussions. 

\newpage


\begin{thebibliography}{999}
%
\setlength{\itemsep}{0.5pt plus1pt minus1pt}
%
\bibitem{BNL} G.\ W.\ Bennett {\it et al.}, Phys.\ Rev.\ {\bf D73}
(2006) 072003.

\bibitem{HMNT03} K.\ Hagiwara, A.\ D.\ Martin, Daisuke Nomura and
  T.\ Teubner, Phys.\ Rev.\ {\bf D69} (2004) 093003; Phys.\ Lett.\ {\bf
    B557} (2003) 69. 

\bibitem{KLOE08} KLOE Collaboration, F.\ Ambrosino {\it et al.},
  Phys.\ Lett.\ {\bf B670} (2009) 285. %, {\tt arXiv:0809.3950}.

\bibitem{KLOE10} 
KLOE Collaboration, F.\ Ambrosino {\it et al.}, 
  %``Measurement of sigma(e+ e- -> pi+ pi-) from threshold to 0.85 GeV^2 using
  %Initial State Radiation with the KLOE detector,''
  Phys.\ Lett.\ {\bf B700} (2011) 102. %, {\tt arXiv:1006.5313v2 [hep-ex]}.

\bibitem{BaBar2pi} BaBar Collaboration, B.\ Aubert {\it et al.},
  Phys.\ Rev.\ Lett. {\bf 103} (2009) 231801.

\bibitem{HMNT06} K.\ Hagiwara, A.\ D.\ Martin, Daisuke Nomura and
  T.\ Teubner, Phys.\ Lett.\ {\bf B649} (2007) 173.

\bibitem{MCSIGHAD} S.\ Actis {\it et al.}, Eur.\ Phys.\ J.\ {\bf C66}
  (2010) 585, {\tt arXiv:0912.0749}.

\bibitem{Ignatov} F.\ Ignatov (for the CMD-2 and SND Collaborations),
  Nucl.\ Phys.\ Proc.\ Suppl. {\bf 181-182} (2008) 101.

\bibitem{KLOE05} KLOE Collaboration, A.\ Aloisio {\it et al.}, Phys.\
  Lett.\ {\bf B606} (2005) 12.

\bibitem{Davier:2010nc}
  M.~Davier, A.~Hoecker, B.~Malaescu and Z.~Zhang,
  %``Reevaluation of the Hadronic Contributions to the Muon g-2 and to
  %alpha(MZ),''
   Eur.\ Phys.\ J.\ {\bf C71} (2011) 1515, {\tt arXiv:1010.4180 [hep-ph]}.

\bibitem{CMD2old} 
  CMD-2 Collaboration, R.\ R.\ Akhmetshin {\it et al.},  
  %``Reanalysis of Hadronic Cross Section Measurements at CMD-2,''
  Phys.\ Lett.\ {\bf B578} (2004) 285, {\tt arXiv:hep-ex/0308008}.

\bibitem{CMD2new} 
  CMD-2 Collaboration, R.\ R.\ Akhmetshin {\it et al.}, 
  %``High-statistics measurement of the pion form factor in the rho-meson energy
  %range with the CMD-2 detector,''
  Phys.\ Lett.\ {\bf B648} (2007) 28, {\tt arXiv:hep-ex/0610021}.

\bibitem{SNDre} SND Collaboration, M.\ N.\ Achasov {\it et al.},
  %``Update of the e+ e- --> pi+ pi- cross section measured by SND detector  in
  %the energy region 400-MeV < s**(1/2) < 1000-MeV,''
  J.\ Exp.\ Theor.\ Phys.\  {\bf 103} (2006) 380 
  [Zh.\ Eksp.\ Teor.\ Fiz.\  {\bf 130} (2006) 437], {\tt arXiv:hep-ex/0605013}.

\bibitem{DavierBaBar} M.~Davier, A.~Hoecker, B.~Malaescu, C.~Z.~Yuan
  and Z.~Zhang, Eur.\ Phys.\ J.\ {\bf C66} (2010)~1.

\bibitem{WolfeMaltman} C.~E.~Wolfe and K.~Maltman,
  %``Consequences of the BaBar $e^+e^- \to pi^+pi^-$ Measurement for the
  %Determination of Model-Dependent $\rho-\omega$ Mixing Effects in
  %$\Pi_{\rho\omega}(m_{\rho}^2)$ and $(g-2)_\mu$,''
  Phys.\ Rev.\ {\bf D83} (2011) 077301, {\tt arXiv:1011.4511}.

\bibitem{Davier:2009ag}
  M.~Davier {\it et al.},
  %``The Discrepancy Between tau and e+e- Spectral Functions Revisited and the
  %Consequences for the Muon Magnetic Anomaly,''
  Eur.\ Phys.\ J.\ {\bf C66} (2010) 127, {\tt arXiv:0906.5443 [hep-ph]}.

\bibitem{Benayoun}
  M.~Benayoun, P.~David, L.~DelBuono and O.~Leitner,
  %``A Global Treatment Of VMD Physics Up To The $\phi$: I. $e^+e^-$
  %Annihilations, Anomalies And Vector Meson Partial Widths,''
  Eur.\ Phys.\ J.\ {\bf C65} (2010) 211, {\tt arXiv:0907.4047
    [hep-ph]};
%``A Global Treatment Of VMD Physics Up To The phi: II. tau Decay and Hadronic
  %Contributions To g-2,''
  Eur.\ Phys.\ J.\ {\bf C68} (2010) 355, {\tt arXiv:0907.5603
    [hep-ph]}; see also 
%``Can VMD improve the estimate of the muon g-2 ?,''
  Chinese Phys.\ {\bf C34} (2010) 698, {\tt arXiv:0912.1248 [hep-ph]}.

\bibitem{JS} F.~Jegerlehner and R.~Szafron,
  %``rho^0-gamma mixing in the neutral channel pion form factor |F_pi|^2 and its
  %role in comparing e^+ e^- with tau spectral functions,'' 
Eur.\ Phys.\ J.\ {\bf C71} (2011) 1632, {\tt arXiv:1101.2872
  [hep-ph]}. 

\bibitem{VEPP-2000} I.\ B.\ Logashenko (on behalf of the CMD-3 and SND
  collaborations), Chinese Phys.\ {\bf C34} (2010) 669. 

%BaBar, 4pi
\bibitem{Babar4pi} BaBar Collaboration, B.\ Aubert {\it et al.},
  Phys.\ Rev.\ {\bf D71} (2005) 052001.

%BaBar, K+K-pi0, K0spiK
\bibitem{BaBar_1} BaBar Collaboration, B.\ Aubert {\it et al.}, Phys.\
  Rev.\ {\bf D77} (2008) 092002. %, {\tt arXiv:0710.4451}

%BaBar, 2(pi+pi-)pi0, 2(pi+pi-)eta, K+K-pi+pi-pi0
\bibitem{BaBar_2} BaBar Collaboration, B.\ Aubert {\it et al.}, Phys.\
  Rev.\ {\bf D76} (2007) 092005.

%BaBar, 2(pi+pi-pi0)
\bibitem{BaBar_3} BaBar Collaboration, B.\ Aubert {\it et al.}, Phys.\
  Rev.\ {\bf D73} (2006) 052003.

\bibitem{HLMNT1} K.\ Hagiwara, R.\ Liao, A.\ D.\ Martin, Daisuke
  Nomura and T.\ Teubner, Chinese Phys.\ {\bf C34} (2010) 728, {\tt
    arXiv:1001.5401 [hep-ph]}. 

\bibitem{DM1} A.\ Cordier {\it et al.}, Phys.\ Lett.\ {\bf 106B}
  (1981) 155.

\bibitem{Lees:2011zi}
  BaBar Collaboration, J.\ P.\ Lees {\it et al.},
  %``Cross Sections for the Reactions $e^+e^- \to K^+ K^- \pi^+\pi^-,\, K^+ K^-
  %\pi^0\pi^0$, and $K^+ K^- K^+ K^-$ Measured Using Initial-State Radiation,''
  {\tt arXiv:1103.3001 [hep-ex]}.

\bibitem{MOR} A.\ D.\ Martin, J.\ Outhwaite and M.\ G.\ Ryskin,
  %``A New determination of the QED coupling alpha (M**2(Z)) lets the Higgs off
  %the hook,''
  Phys.\ Lett.\ {\bf B492} (2000) 69.

\bibitem{BaikovCK} P.~A.~Baikov, K.~G.~Chetyrkin and J.~H.~K\"uhn,
  %``Order $\alpha^4_s$ QCD Corrections to $Z$ and $\tau$ Decays,''
  Phys.\ Rev.\ Lett.\  {\bf 101} (2008) 012002, {\tt arXiv:0801.1821
    [hep-ph]}. 

\bibitem{PDG10} K.\ Nakamura {\it et al.} (Particle Data Group),
  J. Phys. {\bf G37} (2010) 075021.

%CMD-2, K+K-
\bibitem{CMD2_1} CMD-2 Collaboration, R.\ R.\ Akhmetshin {\it et al.},
  Phys.\ Lett.\ {\bf B669} (2008) 217.

%SND, K+K-
\bibitem{SND_1} SND Collaboration, M.\ N.\ Achasov {\it et al.},
  Phys.\ Rev.\ {\bf D76} (2007) 072012.%, {\tt arXiv:0707.2279}

%SND, KsKl
\bibitem{SND_2} SND Collaboration, M.\ N.\ Achasov {\it et al.}, J.\
  Exp.\ Theor.\ Phys. {\bf 103} (2006) 720.%, {\tt arXiv:hep-ex/0606057}

%CMD-2, pi+pi-pi0
\bibitem{CMD2_2} CMD-2 Collaboration, R.\ R.\ Akhmetshin {\it et al.},
  Phys.\ Lett.\ {\bf B642} (2007) 203.

%KLOE, omegapi0
\bibitem{KLOEomegapi} KLOE Collaboration, F.\ Ambrosino {\it et al.},
  Phys.\ Lett.\ {\bf B669} (2008) 223.

%BES
\bibitem{BES_1} BES Collaboration, M.\ Ablikim {\it et al.}, Phys.\
  Rev.\ Lett.\ {\bf 97} (2006) 262001. %, {\tt arXiv:hep-ex/0612054}

%BES-II
\bibitem{BES_2} BES Collaboration, M.\ Ablikim {\it et al.}, Phys.\
  Lett.\ {\bf B677} (2009) 239.

%CLEO
\bibitem{CLEO} CLEO Collaboration, D.\ Besson {\it et al.}, Phys.\
  Rev.\ {\bf D}{\bf 76} (2007) 072008. %, {\tt arXiv:0706.2813}

\bibitem{Harlander:2002ur}
  R.~V.~Harlander and M.~Steinhauser,
  %``rhad: A program for the evaluation of the hadronic R-ratio in the
  %perturbative regime of QCD,''
  Comput.\ Phys.\ Commun.\ {\bf 153} (2003) 244, {\tt arXiv:hep-ph/0212294}.

\bibitem{Aoyama:2008hz}
  T.~Aoyama, M.~Hayakawa, T.~Kinoshita and M.~Nio,
  %``Tenth-Order Lepton Anomalous Magnetic Moment -- Second-Order Vertex
  %Containing Two Vacuum Polarization Subdiagrams, One Within the Other,''
  Phys.\ Rev.\ {\bf D78} (2008) 113006, {\tt arXiv:0810.5208 [hep-ph]}.

\bibitem{Aoyama:2010yt}
  K.~Asano, M.~Hayakawa, T.~Kinoshita, M.~Nio and N.~Watanabe,
  %``Tenth-order lepton g-2: Contribution from diagrams containing a sixth-order
  %light-by-light-scattering subdiagram internally,''
  Phys.\ Rev.\ {\bf D81} (2010) 053009, {\tt arXiv:1001.3704 [hep-ph]}.

\bibitem{Aoyama_latest}
  T.~Aoyama, M.~Hayakawa, T.~Kinoshita and M.~Nio,
  %``Tenth-order lepton g-2: Contribution of some fourth-order radiative
  %corrections to the sixth-order g-2 containing light-by-light-scattering
  %subdiagrams,''
  Phys.\ Rev.\  {\bf D82} (2010) 113004, 
%  [arXiv:1009.3077 []].
%\cite{Aoyama:2011rm}
  %``Tenth-Order QED contribution to Lepton Anomalous Magnetic Moment -
  %Fourth-Order Vertices Containing Sixth-Order Vacuum-Polarization
  %Subdiagrams,''
%  Phys.\ Rev.\  
{\bf D83} (2011) 053002, 
%  [arXiv:1101.0459 []].
%\cite{Aoyama:2010zp}
  %``Proper Eighth-Order Vacuum-Polarization Function and its Contribution to
  %the Tenth-Order Lepton g-2,''
%  Phys.\ Rev.\  
{\bf D83} (2011) 053003.
%  [arXiv:1012.5569 []].

\bibitem{Kinoshitaetal} T.\ Kinoshita and M.\ Nio, Phys.\ Rev.\ {\bf
    D73} (2006) 053007.

\bibitem{Aoyamaetal} T.~Aoyama, M.~Hayakawa, T.~Kinoshita and M.~Nio,
  Phys.\ Rev.\ {\bf D77} (2008) 053012. 

\bibitem{Jegerlehner:2009ry}
  F.~Jegerlehner and A.~Nyffeler,
  %``The Muon g-2,''
  Phys.\ Rept.\ {\bf 477} (2009) 1, {\tt arXiv:0902.3360 [hep-ph]}.

\bibitem{Czarnecki:2002nt}
  A.~Czarnecki, W.~J.~Marciano and A.~Vainshtein,
  %``Refinements in electroweak contributions to the muon anomalous magnetic
  %moment,''
  Phys.\ Rev.\ {\bf D67} (2003) 073006, 
  Erratum-ibid.\ {\bf D73} (2006) 119901, 
  {\tt arXiv:hep-ph/0212229}.

\bibitem{Czarnecki:1995sz}
  A.~Czarnecki, B.~Krause and W.~J.~Marciano,
  %``Electroweak corrections to the muon anomalous magnetic moment,''
  Phys.\ Rev.\ Lett.\ {\bf 76} (1996) 3267, {\tt arXiv:hep-ph/9512369}.

\bibitem{Czarnecki:1995wq}
  A.~Czarnecki, B.~Krause and W.~J.~Marciano,
  %``Electroweak Fermion loop contributions to the muon anomalous magnetic
  %moment,''
  Phys.\ Rev.\ {\bf D52} (1995) 2619, {\tt arXiv:hep-ph/9506256}.

\bibitem{Knecht:2002hr}
  M.~Knecht, S.~Peris, M.~Perrottet and E.~De Rafael,
  %``Electroweak hadronic contributions to g(mu)-2,''
  JHEP {\bf 0211} (2002) 003, {\tt arXiv:hep-ph/0205102}.

\bibitem{Peris:1995bb}
  S.~Peris, M.~Perrottet and E.~de Rafael,
  %``Two--Loop Electroweak Corrections to the Muon g-2: a new class of Hadronic
  %Contributions,''
  Phys.\ Lett.\ {\bf B355} (1995) 523, {\tt arXiv:hep-ph/9505405}.

\bibitem{Prades:2009tw}
  J.~Prades, E.~de Rafael and A.~Vainshtein,
  %``Hadronic Light-by-Light Scattering Contribution to the Muon Anomalous
  %Magnetic Moment,''
  {\tt arXiv:0901.0306 [hep-ph]}.

\bibitem{Nyffeler:2009tw}
  A.~Nyffeler,
  %``Hadronic light-by-light scattering in the muon g-2: a new short-distance
  %constraint on pion-exchange,''
  Phys.\ Rev.\ {\bf D79} (2009) 073012, {\tt arXiv:0901.1172 [hep-ph]}.

\bibitem{Nyffeler:2010rd}
  A.~Nyffeler,
  %``Hadronic light-by-light scattering contribution to the muon g-2,''
  Chinese Phys.\ {\bf C34} (2010) 705, {\tt arXiv:1001.3970 [hep-ph]}.

\bibitem{Erler:2006vu}
  J.~Erler and G.~T.~Sanchez,
  %``An upper bound on the hadronic light-by-light contribution to the muon
  %g-2,''
  Phys.\ Rev.\ Lett.\ {\bf 97} (2006) 161801, {\tt arXiv:hep-ph/0605052}.

\bibitem{Blum:2009zz}
  T.~Blum and S.~Chowdhury,
  %``Hadronic Contributions To G-2 From The Lattice,''
  Nucl.\ Phys.\ Proc.\ Suppl.\ {\bf 189} (2009) 251; S.~Chowdhury {\it
    et al.}, PoS LATTICE2008 (2008) 251.

\bibitem{PRakow} P.\ Rakow, {\em private communications}; see also
  talk at {\em Lattice 2008}, Williamsburg, Virginia, USA, 14--19 July
  2008,\\ {\tt
 http://conferences.jlab.org/lattice2008/talks/parallel/paul\_rakow.pdf}$\,$.

\bibitem{Goecke:2010if}
  T.~Goecke, C.~S.~Fischer and R.~Williams,
  %``Hadronic light-by-light scattering in the muon g-2: a Dyson-Schwinger
  %equation approach,''
  Phys.\ Rev.\ {\bf D83} (2011) 094006,
  {\tt arXiv:1012.3886}.

\bibitem{Boughezal:2011vw}
  R.~Boughezal and K.~Melnikov,
  %``Hadronic light-by-light scattering contribution to the muon magnetic
  %anomaly: constituent quark loops and QCD effects,''
  {\tt arXiv:1104.4510 [hep-ph]}.

\bibitem{Mohr:2008fa}
  P.~J.~Mohr, B.~N.~Taylor and D.~B.~Newell (CODATA 2006),
  %``CODATA Recommended Values of the Fundamental Physical Constants: 2006,''
  Rev.\ Mod.\ Phys.\ {\bf 80} (2008) 633, {\tt arXiv:0801.0028
    [physics.atom-ph]}. 

\bibitem{Roberts:2010cj}
  B.~L.~Roberts,
  %``Status of the Fermilab Muon $(g-2)$ Experiment,''
  Chinese Phys.\ {\bf C34} (2010) 741, {\tt arXiv:1001.2898 [hep-ex]}.

\bibitem{Amsler:2008zzb}
  C.~Amsler {\it et al.} (Particle Data Group),
  %``Review of particle physics,''
  Phys.\ Lett.\ {\bf B667} (2008) 1.

\bibitem{HLMNTprocs2010} K.\ Hagiwara, R.\ Liao, A.\ D.\ Martin,
  Daisuke Nomura and T.\ Teubner, AIP Conf.\ Proc.\ {\bf 1343} (2011)
  340; in the proceedings of {\em Tau2010, 11th International Workshop
    on Tau Lepton Physics, Manchester, U.K., 13--17 September 2010},
  to appear in Nucl. Phys. {\bf B} (Proc.\ Suppl.). 
%in the proceedings. 
%work presented at the conferences {\em QCHS9, Quark Confinement and
%  the Hadron Spectrum IX, Madrid, Spain, 30 August -- 3 September
%  2010} and {\em Tau2010, 11th International Workshop on Tau Lepton
%  Physics, Manchester, U.K., 13--17 September 2010}, to be published
%in the proceedings. 

\bibitem{Passera:2008jk}
  M.~Passera, W.~J.~Marciano and A.~Sirlin,
  %``The Muon g-2 and the bounds on the Higgs boson mass,''
  Phys.\ Rev.\ {\bf D78} (2008) 013009, {\tt arXiv:0804.1142 [hep-ph]}.

\bibitem{Passera:2010ev}
  M.~Passera, W.~J.~Marciano and A.~Sirlin,
  %``The muon g-2 discrepancy: new physics or a relatively light Higgs?,''
  Chinese Phys.\ {\bf C34} (2010) 735, {\tt arXiv:1001.4528 [hep-ph]}.

\bibitem{Steinhauser:1998rq}
  M.~Steinhauser,
  %``Leptonic contribution to the effective electromagnetic coupling constant up
  %to three loops,''
  Phys.\ Lett.\ {\bf B429} (1998) 158.

\bibitem{Delaltop}
 K.~G.~Chetyrkin, J.~H.~K\"uhn and M.~Steinhauser,
 Phys.\ Lett.\ {\bf B371} (1996) 93;
 Nucl.\ Phys.\ {\bf B482} (1996) 213;
 Nucl.\ Phys.\ {\bf B505} (1997) 40.

\bibitem{Kuhn:1998ze}
  J.~H.~K\"uhn and M.~Steinhauser,
  %``A theory driven analysis of the effective QED coupling at M(Z),''
  Phys.\ Lett.\ {\bf B437} (1998) 425, {\tt arXiv:hep-ph/9802241}.

\bibitem{de Troconiz:2004tr}
  J.~F.~de Troconiz and F.~J.~Yndurain,
  %``The hadronic contributions to the anomalous magnetic moment of the  muon,''
  Phys.\ Rev.\ {\bf D71} (2005) 073008, {\tt arXiv:hep-ph/0402285}.

\bibitem{Burkhardt:2005se}
  H.~Burkhardt and B.~Pietrzyk,
  %``Low energy hadronic contribution to the QED vacuum polarization,''
  Phys.\ Rev.\ {\bf D72} (2005) 057501, {\tt arXiv:hep-ph/0506323}.

\bibitem{Jegerlehner:2008rs}
  F.~Jegerlehner,
  %``The running fine structure constant alpha(E) via the Adler function,''
  Nucl.\ Phys.\ Proc.\ Suppl.\ {\bf 181-182} (2008) 135, {\tt
    arXiv:0807.4206 [hep-ph]}. 

\bibitem{FJ2010}
  F.~Jegerlehner, talk presented at the {\em LC10 Workshop, INFN/LNF
    Frascati, Italy, December 2010}, {\tt arXiv:1107.4683 [hep-ph]}.

\bibitem{EWWG} For latest results of the `LEP Electroweak Working
  Group' (LEP EWWG) see {\tt http://lepewwg.web.cern.ch/LEPEWWG/}. 

\bibitem{MGrunewald} M.~Gr\"unewald, {\em private communications}. For
  a description of the procedure used see: ALEPH Collaboration, {\tt
    arXiv:1012.2367 [hep-ex]}. 

\bibitem{HLMNT2} K.\ Hagiwara, R.\ Liao, A.\ D.\ Martin, Daisuke
  Nomura and T.\ Teubner, in preparation.

\bibitem{ChoHagiwaraMatsumotoNomura}  G.~C.~Cho, K.~Hagiwara,
  Y.~Matsumoto and D.~Nomura, 
  %``The MSSM confronts the precision electroweak data and the muon g-2,''
  {\tt arXiv:1104.1769}. 

\end{thebibliography}
\end{document}